\documentclass[10pt,journal,compsoc]{IEEEtran}
\usepackage[numbers]{natbib}
\usepackage{comment, multirow, enumitem, color, yfonts, graphicx}
\usepackage{algorithm, amsmath, gensymb, amssymb, mathtools}  
\usepackage{enumitem}
\usepackage{pifont}
\usepackage[normalem]{ulem}
\usepackage{hyperref}
\usepackage{fancyhdr} %
\usepackage{pdfpages}
\usepackage{enumitem}
\usepackage{multirow, subfigure, url, caption}
\usepackage{algorithmicx, algpseudocode, xspace}
\usepackage{tikz, booktabs}
\usepackage{xcolor}
\usepackage{listings}
\usepackage{inputenc}

\definecolor{mGreen}{rgb}{0,0.6,0}
\definecolor{midnightblue}{rgb}{0,0.2,0.4}
\definecolor{mGray}{rgb}{0.5,0.5,0.5}
\definecolor{mPurple}{rgb}{0.58,0,0.82}
\definecolor{backgroundColour}{rgb}{0.95,0.95,0.92}

\lstdefinestyle{CStyle}{
	backgroundcolor=\color{backgroundColour},   
	commentstyle=\color{midnightblue},
	keywordstyle=\color{blue},
	numberstyle=\scriptsize\color{black},
	stringstyle=\color{black},
	basicstyle=\footnotesize,
	breakatwhitespace=false,         
	breaklines=true,                 
	captionpos=b,                    
	keepspaces=true,                 
	numbers=left,                    
	numbersep=-8pt,                  
	showspaces=false,                
	showstringspaces=false,
	showtabs=false,                  
	tabsize=3,
	language=C	
}

\makeatletter
\algrenewcommand\ALG@beginalgorithmic{\small}
\newcommand{\algmargin}{\the\ALG@thistlm}
\makeatother
\algrenewcommand{\algorithmiccomment}[1]{\footnotesize \ $//$#1 \small}
\algnewcommand{\parState}[1]{\State%
	\parbox[t]{\dimexpr\linewidth-\algmargin}{\strut #1\strut}}

\newcommand{\iRobot}{iRobot\ }
\newcommand{\setequationmargin}{\setlength{\belowdisplayskip}{3pt}\setlength{\belowdisplayshortskip}{3pt}\setlength{\abovedisplayskip}{3pt}\setlength{\abovedisplayshortskip}{3pt}}

\newcommand*\circled[1]{\tikz[baseline=(char.base)]{
		\node[shape=circle,draw,inner sep=0.5pt] (char) {#1};}}

\usepackage[listings,skins]{tcolorbox}
\usepackage{color}

\newcommand{\RT}{\textsc{Robo\-Fuzz}\xspace}

\ifCLASSOPTIONcompsoc
\else
\fi
\ifCLASSINFOpdf
\else
\fi
\begin{document}
%
\title{Securing Autonomous Service Robots through Fuzzing, Detection, and Mitigation}
%
%
%
%

\author{Chundong~Wang,
        Yee~Ching~Tok,
        Rohini~Poolat,
        ~Sudipta~Chattopadhyay,
        and Mohan Rajesh Elara
\IEEEcompsocitemizethanks{\IEEEcompsocthanksitem C. Wang, Y. C. Tok, S. Chattopadhyay, and M. R. Elara
	are with Singapore Universtiy of Technology and Design, Singapore
	\protect\\
E-mail: cd\_wang@outlook.com, yeeching\_tok@mymail.sutd.edu.sg, sudipta\_chattopadhyay@sutd.edu.sg, and rajeshelara@sutd.edu.sg.
\IEEEcompsocthanksitem R. Poolat is with National University of Singapore. This work was done when she worked in Singapore University of Technology and Design.
	\protect \\
Email: rohini\_poolat@yahoo.com.
}
\thanks{Manuscript received MM DD, 2020; revised MM DD, 2020.}}

%
%

\markboth{IEEE Tranasctions on XXXXXX}%
{On Securing Autonomous Service Robots through Fuzzing, Detection and Mitigation}
%



\IEEEtitleabstractindextext{%
\begin{abstract}
Autonomous service robots share social spaces with humans, usually working together for domestic or 
professional tasks. 
Cyber security breaches in such robots undermine the trust between humans 
and robots. In this paper, we investigate how to apprehend and inflict security threats at the
design and implementation stage of a movable autonomous service robot.
To this end, we leverage the idea of directed fuzzing and design \RT that systematically
tests an autonomous service robot in line with the robot's states 
and the surrounding environment. The methodology of \RT is to study critical environmental
parameters affecting the robot's state transitions and
subject the robot control program with 
rational but harmful sensor values so as to compromise the robot.
Furthermore, we develop detection and mitigation
algorithms to counteract the impact of \RT. The difficulties mainly lie in
the trade-off among 
limited computation resources, timely detection and
the retention of work efficiency in mitigation. 
In particular,
we propose detection and mitigation methods that take advantage of historical records 
of obstacles to detect inconsistent obstacle appearances regarding untrustworthy sensor values
and navigate the movable robot to continue moving so as to carry on a planned task.
By doing so, we manage to maintain a low cost for detection and mitigation but also
retain the robot's work efficacy.
We have prototyped the bundle of \RT, detection and mitigation algorithms
in a real-world movable robot. Experimental results confirm that
\RT makes a success rate of 
up to 93.3\% in imposing concrete threats to the robot while
the overall loss of work efficacy is merely 4.1\% at the mitigation mode.

\end{abstract}

}

\maketitle

\IEEEdisplaynontitleabstractindextext

%
\IEEEpeerreviewmaketitle

\section{Introduction}

Autonomous service robots are widely used to  
not only relieve people from dirty, 
monotonous, and dull tasks, but also reduce economic costs~\cite{Fiorini2000, 1242021, robot:service:RAS-2013, robot:service:RAS-2017, 8793764}. 
For example, cleaning robots gain wide popularity in
tidying private apartments and public places. 
 Seoul-Incheon and Singapore Changi airports have deployed
cleaning robots to replace human cleaners and the latter should 
save 20\% housekeeping manpowers~\cite{robot:LG:2017, robot:changi-T4:Aug-2017}.
Since autonomous service robots are sharing social spaces with humans at home, 
in the offices and even in critical infrastructures like
airports and banks, their security and safety are of paramount importance, 
especially concerning they are autonomous without human attendance.

Robotics are generally categorized as 
cyber-physical systems (CPS).
A robot
typically has
1) a digital controller, e.g., Raspberry Pi, to manage the system,  
2) physical components, such as sensors and actuators, to sense the surrounding
environment (e.g., distance) and to manipulate physical entities (e.g., wheels and 
robotic arms), respectively, and 3)
cyber components that connect the robot to  
networks (e.g., for remote control via smartphones).
The robot control program running in the controller 
is critical to the security and safety of a robot as it decides how to
manoeuvre actuators of the robot on reading sensor values. 
A number of studies have 
revealed that it is possible
to compromise a CPS through fraudulent sensor values, while 
mitigating such attacks usually requires  
the involvement of a cloud server for remote computation or attestation~\cite{security:power-grid:CCS-2009,security:stealthy-attacks:CCS-2016,security:CPS:DATE-2017,security:CPS-sensor:DAC-2017,security:Orpheus:ACSAC-2017}.
However, such methods are not applicable to autonomous service robot. The reason is threefold.
Firstly, the computational resource and battery capacity 
are relatively limited for an economical autonomous service robot compared to large CPS, say, a power grid.
Secondly, the vast popularity of autonomous service robots imposes overwhelming difficulty
on security patches or remote attestation from time to time. 
Thirdly, many autonomous service robots move themselves to 
complete planned tasks, which differentiates them from stationary CPS like power grid or 3D printer and
necessitates a mitigation method that replenishes the movement of autonomous service robot.
As a result, it is preferable and practical to secure an autonomous service robot as early as at its design and implementation stage.

In this paper, we work at the standpoint of developers 
to enhance the security and safety 
of autonomous service robots, particularly ones that are movable
because they would be physically detrimental to human beings once compromised.
We would proceed at two dimensions. On one hand, 
we attempt to systematically scrutinize
the security threats to autonomous service robot through investigating 
the values of critical sensors, since
these sensor values, as inputs to the robot control program, determine 
the next states of robot. On the other hand, with regard to the uncovered threats,
we develop an efficient algorithm to mitigate their impacts while retaining most of 
the robot's work efficacy.

Without loss of generality, we illustrate with
an autonomous service robot cruising 
by means of an ultrasonic distance sensor to avoid obstacles. 
Once the distance sensor indicates a close obstacle ahead, 
the robot control program ought to direct the robot to turn left or right. 
Otherwise, the robot would
crash into the obstacle. As a result, 
altering the sensor value to be malicious for the robot control program
would inflict serious threats to the robot. 
We hence employ the idea of software {\em fuzzing} to test the robot control program. 
In software testing, fuzzing is used to   
identify security vulnerabilities or bugs in a program
by subjecting the program to various kinds of input,
and the program may crash or yield absurd outputs~\cite{fuzzing:book-2008, fuzz:radamsa, security:fuzzing:CCS-2017}.
By fuzzing the robot control program, we aim to discover as many flaws as possible 
in the robot control program and secure the robot in turn.

We use a state-of-the-art fuzzing tool, 
i.e., Radamsa~\cite{fuzz:radamsa},
to fuzz the robot control program 
of the aforementioned movable robot employing a distance sensor 
for motion. We generate and feed a series 
of distance sensor values
to the robot control program
replacing real-world distances when the robot is moving. 
The robot trembles because the sensor values fuzzed by Radamsa,  
as intended to maximally uncover 
bugs of a program, fluctuate significantly.  
We can easily patch the robot control program with a filter
to rule out such volatile sensor values, since
a moving robot working in a rational environment is 
expecting sensor values that fall in a reasonable
range in line with the environment and the state of robot. 
For instance, a robot moving towards a wall should
continuously receive decreasing distance sensor values.

The analysis over arbitrary and irregular sensor values, 
however, implies us further in fuzzing the robot control program, i.e.,
to test the control program with rational and regular sensor values
in a {\em directed fuzzing} way. The distance can be viewed as a critical 
{\em environmental parameter} for a moving service robot as it 
triggers state transitions for the robot.
For example,
a moving robot receiving decreasing distance sensor values 
would turn left or right while the distance gradually drops 
below a threshold.
Given a dynamic obstacle, say,
an automatic sliding door, it may move out of the robot's path and 
the distance sensor value should
suddenly increase to a large value, after which the robot 
keeps moving forward. 
Note that the robot control program is unable to ascertain whether an obstacle
is truly dynamic or static solely depending on distance sensor values,
because the scenarios where 
the distance either monotonically decreases or abruptly increases
are both possible in the real world.
Assume that the robot is moving towards a hard wall, but 
we intentionally replace the distance sensor values with ones that resemble
the getaway of a dynamic obstacle. The robot should 
collide with the wall.

The aforementioned example addresses the essence of our directed fuzzing strategy, namely \RT.
In a nutshell, by 
investigating the state transitions and environmental parameters that
influence the behaviors of an autonomous service robot,  
\RT generates rational but harmful sensor values so as to mislead the robot 
for concrete threats.

Adversaries can implement 
\RT with realistic attack models, 
like suspending or fabricating sensor values,
to compromise an autonomous service robot.
As developers, we move forward and 
defend against the attacks entailed via \RT
by detecting and mitigating them. There are two concerns in doing so.
First, the detection and mitigation should not be
heavyweight regarding the limited computational resources of an autonomous
service robot.
Secondly, once an attack is detected, the mitigation cannot barely shut down the 
robot but maximally retain the robot's work efficacy. Nevertheless,
as mentioned, the robot control program alone  
cannot rule out rational but anomalous sensor values.
We need further information that can be used to counteract \RT.
We note that, for a movable service robot, such as a cleaning robot,
it is supposed to repeatedly cruise in a certain and steady place. 
Consequently, the robot is able to make and maintain a historical record of 
obstacles for the place~\cite{robot:roomba-map:2018}. Such a 
historical record is an exploitable resource for us to detect the attacks initiated through 
\RT. Concerning that
\RT deceives the robot control program using fuzzed distance sensor values that emulate a dynamic obstacle,
a historical record can help the robot control program to cross-check 
if the obstacle is really dynamic so as to avoid a collision where necessary.

The historical record is also effectual for us to mitigate the impact caused by \RT.
A movable autonomous robot must keep moving to complete 
the task planned for it even in the 
presence of an untrustworthy distance sensor. 
Although the robot control program cannot put reliance on distance sensor values, 
it can reuse the historical record to 
circumvent obstacles and
navigate the robot. Reusing such records not only retains a movable robot's work efficacy,
but also gains high cost efficiency in mitigating for an economical robot.

The main ideas of this paper are summarized as follows.
\begin{itemize}[leftmargin=5mm,nosep]
	\setlength{\itemsep}{-\itemsep}	
	\item We propose \RT which tests an autonomous service robot by fuzzing rational but harmful sensor values so as to mislead the robot's control program;
	\item To defend against the attacks initiated by \RT, we develop detection and mitigation methods which leverage historical records to maximally protect the robot and efficiently accomplish planned tasks.
\end{itemize}
\RT and strategies of detection and mitigation to contract \RT
form a self-contained and systematic scheme that 
help to develop a 
secure autonomous service robot. We have prototyped them with a real-world movable robot,
i.e., iRobot Create 2 with an HC-SR04 distance sensor. 
Experimental results confirm that \RT attains
up to 93.3\% success rate in imposing threats  
onto a moving iRobot Create 2.
Our detection and mitigation methods also  
efficiently detect attacks at a very high rate and make the
robot being under attack accomplish scheduled tasks with an
insignificant loss of work efficacy, i.e., 4.1\% overall.

The remainder of this paper is organized as follows.
In Section~\ref{sec:bg}, we present the background of autonomous service robot
and attack models it is prone to.
We conduct a motivational study to incur concrete threats to a movable service robot
in Section~\ref{sec:mot}. In Section~\ref{sec:fuzzing}, we detail the design 
and methodology of \RT. In Section~\ref{sec:detection} and Section~\ref{sec:migitation}, respectively,
we show our algorithms for detecting and mitigating threats incurred by \RT.
We present experimental results with a prototype built with iRobot Create 2 in Section~\ref{sec:evaluation}.
We discuss threats to validity in Section~\ref{sec:threats} and
conclude the paper in Section~\ref{sec:conclusion}.

\section{Background and Related Work}~\label{sec:bg}

In contrast to robots used by manufacturers or specialists, 
service robots are close 
to people and easy to operate, 
providing a variety of services, 
such as housekeeping and entertainment~\cite{robot:service:RAS-2013, robot:service:RAS-2017}.
According to ISO standard~\cite{robot:ISO-standard:2012}, 
a service robot is a class 
of robots that ``perform useful tasks for humans or equipment 
excluding industrial automation applications''.
A movable autonomous service robot, such as the typical cleaning robot, 
has following components: 1) a digital controller
such as Raspberry Pi or Arduino Mega where a control program runs, 
2) numerous sensors to sense surroundings, 3) wheels
to move the robot around, and 4) cyber accessories for network connection.
Autonomous service robot puts reliance on the
control program to decide the next move of it in accordance
with sensor values obtained from time to time.
Regarding 
sensors installed in a service robot, 
they quantitatively measure and report the environmental parameters
the robot is encountering. For example, a distance sensor 
tells whether the robot is too close to any obstacle.
Sensors may work in different modes. A sensor working in the proactive
mode alerts the robot control program periodically or in case of emergency
while a sensor working in the passive mode pends the robot control program
to ask for sensor value.

\begin{figure}[t]
	\centering
	\includegraphics[width=\columnwidth]{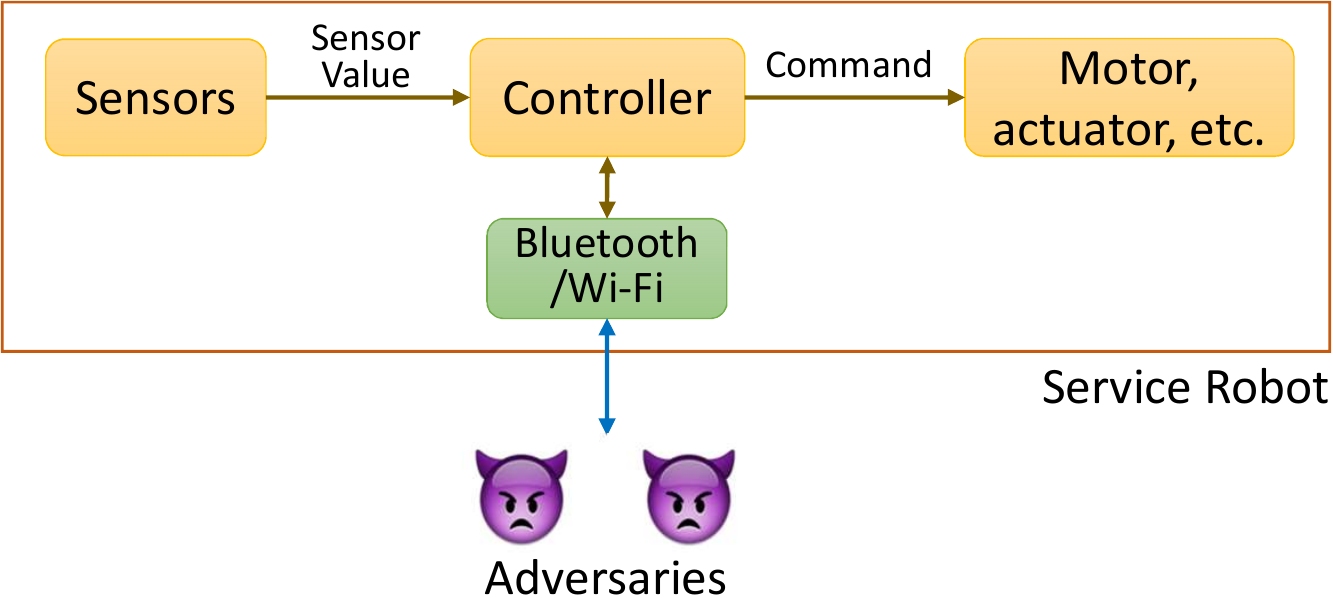}
	\caption{A Illustration of Autonomous Service Robot and Adversaries}~\label{fig:arch}	
\end{figure}

Robots fall into the broad category of
CPS.
One outstanding characteristic of CPS is the vast heterogeneity
of building blocks in different CPS for different usages~\cite{security:CPS:DATE-2017,  security:CPS-survey:IEEE-2017, CPS:R2U2:Spinger-2017, CPS:survey:Springer-2018}.
An autonomous service robot is significantly distinct from 
typical CPS such as power grids, handheld smartphones or 3D printer~\cite{security:smart-home:ICCAD-2014, security:CPS-fuzzing:DAC-2014, security:smart-grid:ICCAD-2015}.   
First, an autonomous service robot is generally a simple system 
with an economical micro-controller and a few hardware components including
sensors, actuators, and network modules.~\autoref{fig:arch} shows a classic architecture of 
autonomous service robot.
Secondly, autonomous service robots gain worldwide popularity in our daily life. 
For example, \iRobot has sold more than 20 million cleaning robots 
since its foundation~\cite{robot:irobot-sale:2018}
 while the sales volume of Xiaomi Mi robots has reached 
 one million in 18 months since its release date~\cite{robot:Mi-sale:2018}. 
Assuming that a critical flaw of cleaning robot 
is uncovered, a large population of users would 
be affected. 
Thirdly, unlike CPS that undergo frequent  
maintenance services
in subways, hospitals, and power stations~\cite{robot:surgical-robot:2015, security:smart-grid:ICCAD-2015, security:fuzzing-transportation:DAC-2015},
many  
service robots are unlikely to be promptly 
upgraded with security patches.
To update a large number of robots or do remote attestation for each
of them is also challenging
and costly for a manufacturer. 
Finally,
a movable service robot is not stationary like 3D printer or handheld smartphone.
Once compromised, it might be manipluated 
to incur physical damages to surrounding people.

\vspace{2ex}
In summary, the demand to study security-related issues 
for autonomous service robots is actual and critical.
Recently, researchers have looked into the cyber security of 
robots~\cite{robot:service:RHIC-2010, robot:security:IntechOpen-2017, SABALIAUSKAITE2017174, robot:hazards:CoRR-2018}.
The security issues should be
considered in the design phase of a service robot due to the ever-increasing 
popularity of service robots and the ever-growing strengths of adversaries. 
In this paper, we first proceed at the standpoint of
developers to explore how to reveal as many flaws as possible
for an autonomous service robot. 
Then we continue to contemplate cost-efficient methods for detection
and mitigation while retaining the robot's work efficacy.

\section{Motivation and Overview}\label{sec:mot}

\begin{figure}[t]
	\centering
	\subfigure[Suspension Attack]{\includegraphics[width=0.8\columnwidth]{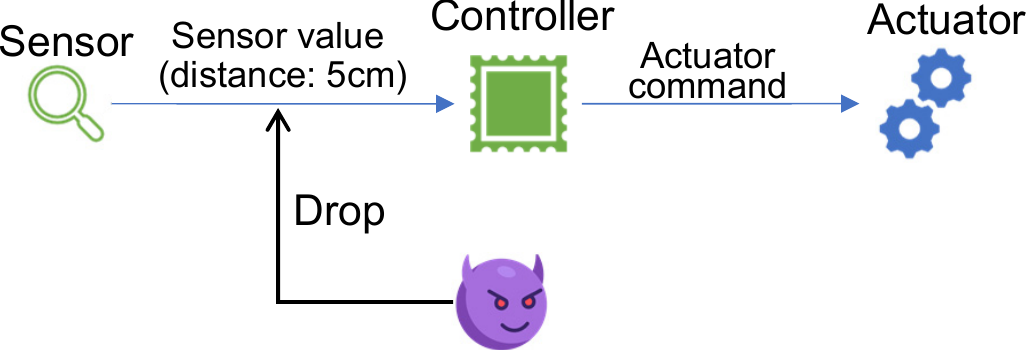}~\label{fig:suspension-model}}
	\subfigure[Fabrication Attack]{\includegraphics[width=0.8\columnwidth]{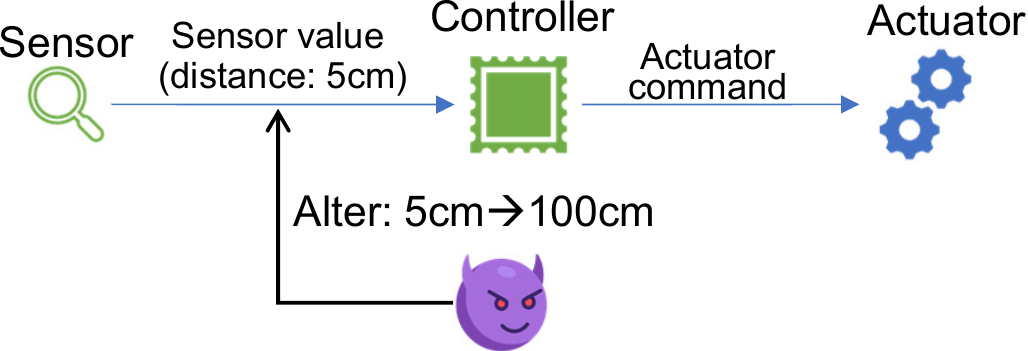}~\label{fig:fabrication-model}}
	\caption{An Illustration of Typical Attacks Models}\label{fig:models}
\end{figure}

\subsection{Security Treats for Autonomous Service Robot}\label{sec:model}

\begin{figure*}[t]
	\centering
	\subfigure[A static obstacle (e.g., a wall)]{\includegraphics[width=0.49\columnwidth]{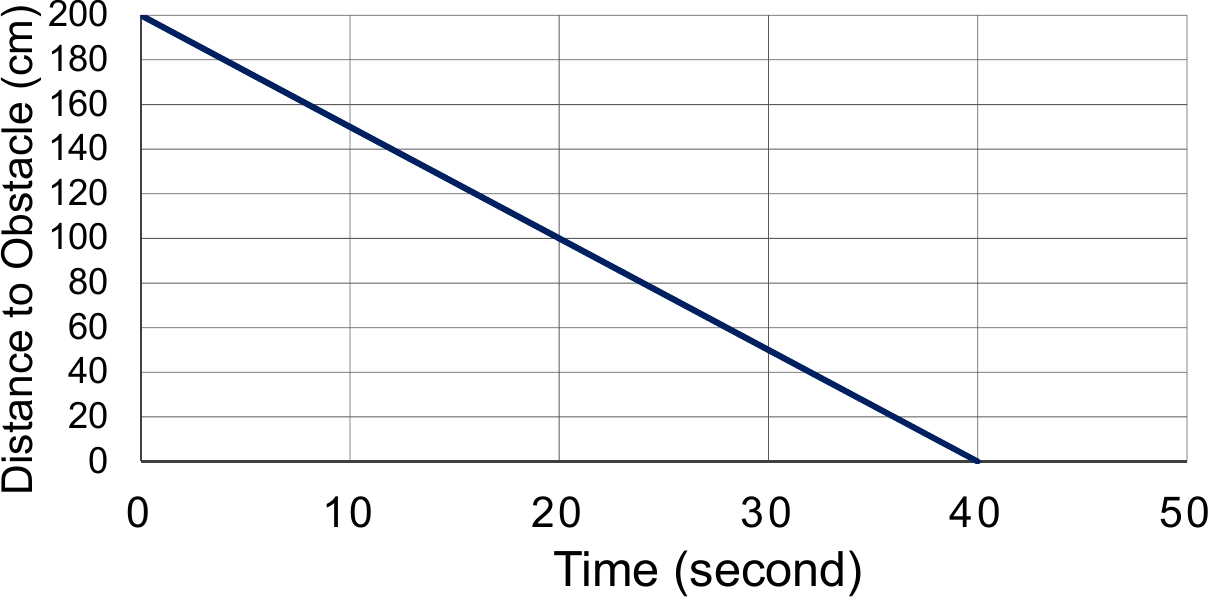}~\label{fig:wall}}
	\hfill
	\subfigure[A dynamic obstacle suddenly moves in the same direction as  robot]{\includegraphics[width=0.49\columnwidth]{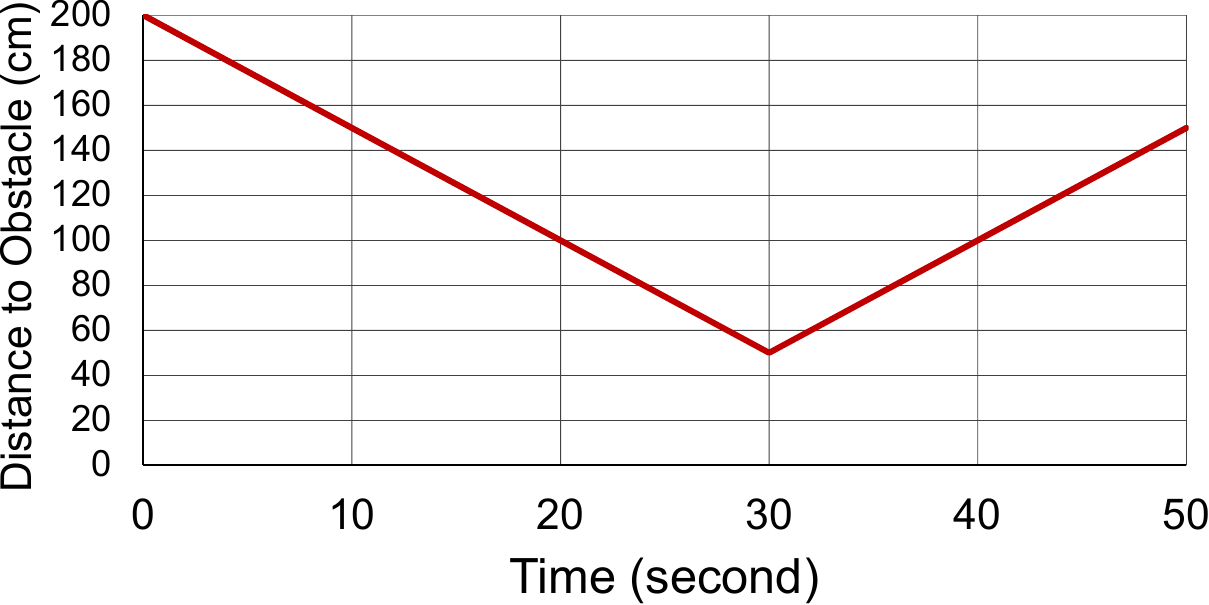}~\label{fig:away}}
	\hfill
	\subfigure[A dynamic obstacle suddenly moves towards the robot]{\includegraphics[width=0.49\columnwidth]{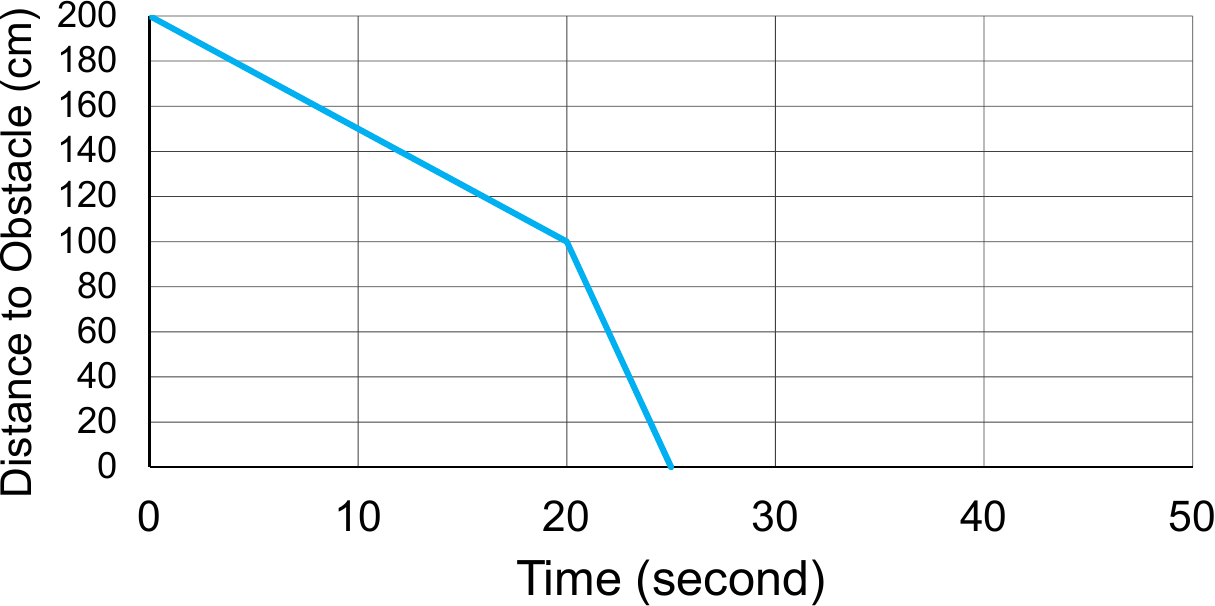}~\label{fig:face}}
	\hfill
	\subfigure[A dynamic obstacle suddenly moves out of the path of robot]{\includegraphics[width=0.49\columnwidth]{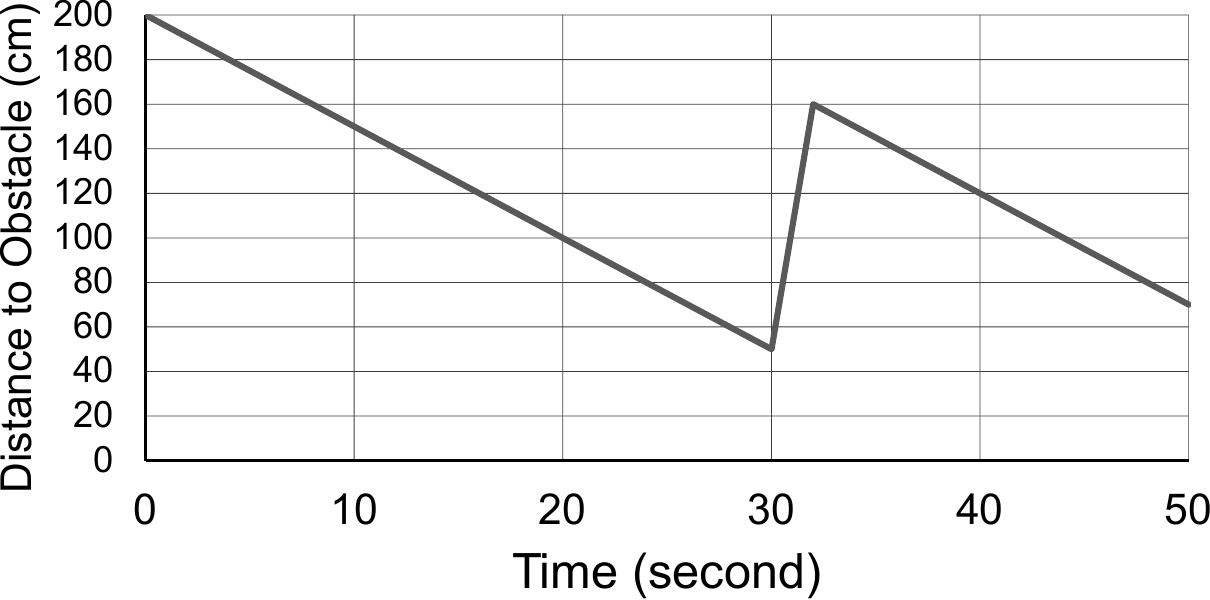}~\label{fig:slide}}\\
	\caption{An Illustration of State Transitions for a Cleaning Robot Regarding Four Types of Obstacles}	\label{fig:states}		
\end{figure*}

Recently, Bonaci et al.~\cite{robot:surgical-robot:2015} investigated the 
vulnerabilities of teleoperated surgical robots and 
Quarta et al.~\cite{robot:industrial-security:SP-2017} performed an empirical 
analysis on the security issues of industrial robots. These works draw 
the attention of research community to the security of robots found in 
factories, operating rooms, and so on. Nevertheless, such awareness 
should be extended to the security of autonomous service robots.
In practice, Giese and Wegemer have managed to hack a Xiaomi 
Mi cleaning robot~\cite{robot:Mi-hack:2018}. Their 
success should not only alert robot manufacturers, but also the users of such 
service robots.

As mentioned, the robot control program maneuvers 
an autonomous service robot by reading sensor values. 
On the other hand, the network interface of robot widely
provides adversaries an exploitable attack surface,
because many users still use default or weak passwords 
today, especially for domestic robots. As a result, 
adversaries are bound to manipulate the robot's sensor values
through unauthorized 
remote access so as to misguide the robot control program.
In the meantime, there are multiple attack models 
for adversaries to follow.
In this paper we consider two
harmful and representative ones that have been manifested recently~\cite{fingerprint:CIDS:Security-2016, SABALIAUSKAITE2017174},
i.e., {\em suspension attack} and {\em fabrication attack}.

\paragraph*{\textbf{Suspension Attack}} 
As shown in Figure \ref{fig:suspension-model}, attackers
suspend sensors from sending out information. 
A sensor at the passive mode, once suspended, would
leave a null response  
to the robot control program, which 
misleads the control program to 
conclude that the sensor malfunctions. 
If the sensor works at the proactive mode,
the impact of suspension attack
should be even worse.
Consider a distance sensor that alerts the robot control program
only in case of a very close obstacle.
After a successful suspension attack,  
the control program would no longer receive any alert. 
As a consequence,
the robot might crash into an obstacle.

\paragraph*{\textbf{Fabrication Attack}} 
With a fabrication attack,  
adversaries fabricate harmful sensor values and feed them to 
the robot control program. 
As shown by Figure~\ref{fig:fabrication-model}, when the robot
is in motion, the control program asks for a distance sensor value
to decide whether an obstacle is nearby to the robot.
Noticing such a request, adversaries replace the normal 
sensor value with an anomalous one. The control
program would accordingly make a wrong decision and put
the robot into a concrete danger.

\subsection{A Motional Study}

Without loss of generality, we choose a programmable 
cleaning robot, i.e., iRobot Create 2~\cite{robot:irobot-create2:2015},
for case study.
We run a control program in 
a Raspberry Pi 3 to maneuver the robot and install an ultrasonic distance sensor (HC-SR04)
to enable the robot to avoid obstacles. 
As developers, we use the WiFi interface shown in~Figure~\ref{fig:arch}
 as the port to communicate with the robot controller
for monitoring and debugging.

The distance to obstacles is 
a crucial environmental parameter
for a cleaning robot.
The control program depends on the distance sensor 
values received at runtime to decide whether the robot moves forward or turns. 
As these sensor values are the input to the control program, 
the first attempt we did  
is leverage the idea of software fuzzing, which
generates various kinds of input values to a program so as to inflict 
disorder or even crash to the program.
We used Radamsa~\cite{fuzz:radamsa}, a
state-of-the-art fuzzing tool,
to make a series of 1,006 values 
 within the distance range supported by HC-SR04 (2cm to 400cm).
A segment of the values fuzzed by Radamsa are as follows:
\setlength{\abovedisplayskip}{3pt}
\setlength{\belowdisplayskip}{3pt}
\begin{equation*}
\{..., 26, 128, 5, 16, 3, 241, 107, 6, 255, 45, 240, 4, 18, ... \}
\end{equation*}
We supposed that such distance sensor values, when fed to 
the control program, should have compromised the robot.
However, after we delivered them  
to satisfy the requests raised by the control program,
the control program would refuse them as
anomalies. 
We then analyzed the failure of fuzzing control program in isolation. 
The reason is mainly due to the 
concept of software fuzzing and the mechanism  
of service robot. 
Fuzzing a program is used to  
reveal bugs and security vulnerabilities
of a program. Hence the fuzzed inputs, as shown in the above segment, 
fluctuate significantly so as to 
traverse different code paths and generate 
as many corner cases as possible. Therefore,
fuzzing the control program is a good  
approach to test the program alone but ignores the
mechanism of service robot. As mentioned, the control program transits a
cleaning robot among states depending on sensor values it receives.
A cleaning robot moving towards a wall will
receive decreasing sensor values and in the end it should
turn or stop, so the robot transits from a state of
moving forward to the next state of turning or stopping.
Given sensor values fuzzed by Radamsa that change strikingly
and continually, they are easy to be distinguished
since they obviously deviate from what the control program expects
in an ordinary environment.

We thoroughly investigate  
the states of cleaning robot
and environmental parameters that drive the robot to do 
state transitions. We find that,
for a cleaning robot
moving at a stable velocity (e.g., 5cm/s), 
its state transitions are affected by the distance to obstacles in 
four scenarios, as ideally 
illustrated by Figure~\ref{fig:states}.
In the four diagrams of Figure~\ref{fig:states},
the Y axis is the distance to obstacles measured over time (cf. X axis).
In Figure~\ref{fig:wall}, the robot is moving towards a fixed
obstacle (e.g., a wall), so the distance gradually
decreases to zero. The remaining three diagrams show
a robot meets three types of dynamic obstacles.
In Figure~\ref{fig:away}, at a time, a dynamic obstacle (e.g., a pet) 
suddenly moves away at a higher velocity and in the same direction as the robot,
so the distance stops dropping but increases abruptly.
In Figure~\ref{fig:face}, after 20s, the dynamic obstacle moves towards
the robot, which makes their distance decrease faster than before.
Figure~\ref{fig:slide} represents another kind of obstacle that
 has been on the path of the robot but, at one moment, moves out of 
 the robot's path, like the prompt open of an automatic sliding door. 
The distance thus migrates to another decreasing linear curve.

The four cases capture normal scenarios where 
the distance to obstacles, as a critical environmental parameter,   
affects a cleaning robot in transiting its states at runtime, say, 
to keep moving forward or turn/stop.
The four curves in Figure~\ref{fig:states} help the control program
rule out anomalous sensor values like ones generated by Radamsa.
More important, they inspire us with the
opportunities to {\em mislead} the control program.
Note that the control program 
relies on the distance sensor values to learn the distance to obstacles.
Consider a cleaning robot is steadily moving to a wall.
We are monitoring the robot's 
state and the real distance by reading sensor values. 
When the robot is close to the wall, 
we fuzz  
increasing distance sensor values to emulate that 
the obstacle is dynamic 
and moving away. If the control program asks
for distance sensor values, we 
will feed fuzzed values to it. 
From the viewpoint
of control program, such increasing
sensor values are absolutely rational regarding Figure~\ref{fig:away}.
So the robot is misled from the curve in Figure~\ref{fig:wall}
to the one in Figure~\ref{fig:away}. In the end, the robot shall crash into the
wall. 

\begin{figure}[t]	
	\centering
	\scalebox{1.00}{\includegraphics[width=\columnwidth]{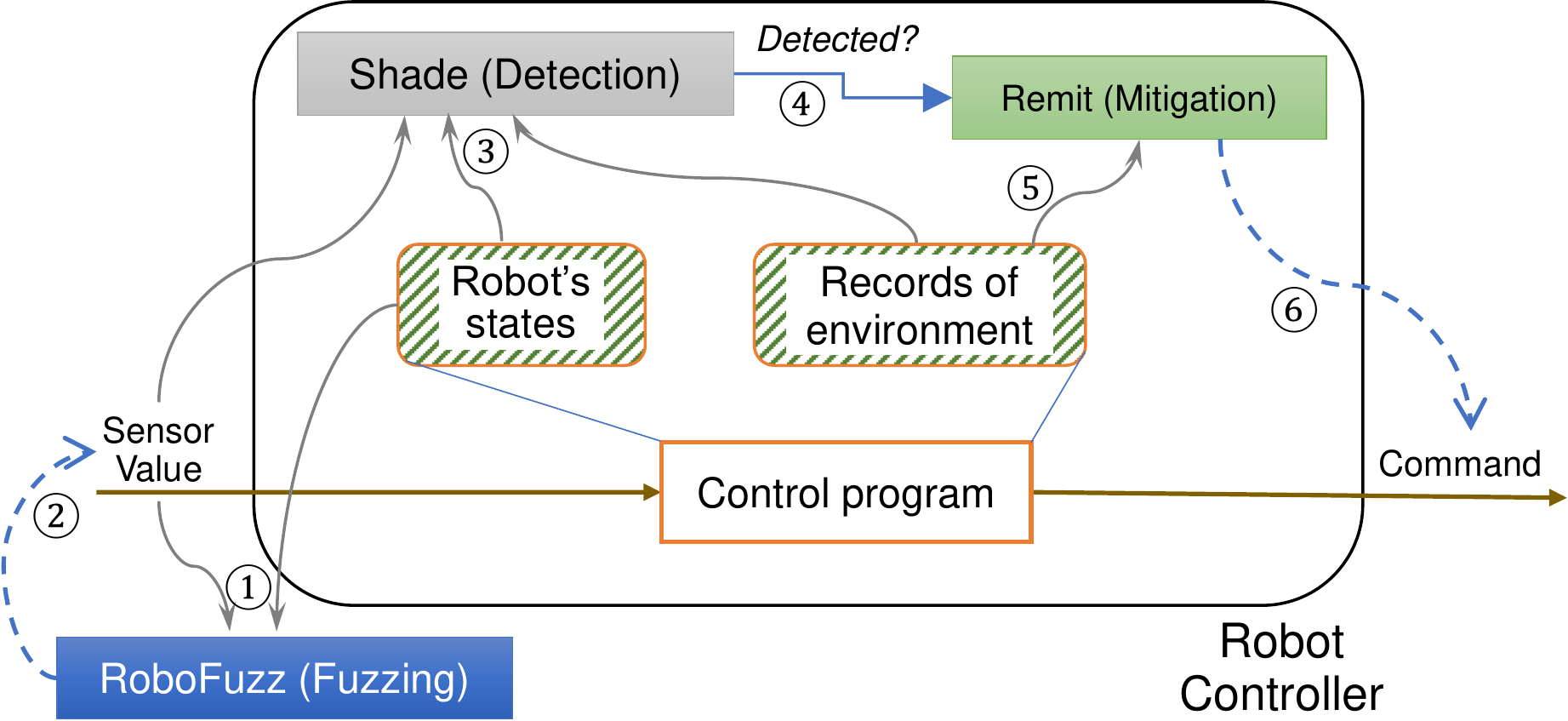}}  
	\caption{An Overview of \RT, Detection and Mitigation}~\label{fig:system}
\end{figure}

\subsection{Overview}

\autoref{fig:system} illustrates an overview of the three schemes proposed in this paper.
The preceding motivating example indicates the essence of \RT (at the leftmost corner 
of~\autoref{fig:system}). 
By closely monitoring the state of a robot and its environmental parameters (\circled{1} in~\autoref{fig:system}), 
\RT starts to deceive the robot's control program at an appropriate occasion with faked but
 rational sensor values ($\circled{2}$ in~\autoref{fig:system}) so as to inflict concrete harm to the movable robot.

The sensor values fuzzed by \RT should impose concrete security breaches to 
autonomous service robots. Because our intention is
to enhance the security and safety of autonomous service robots at their design 
and implementation stage, we need to defend against \RT.  
We subsequently develop detection and mitigation schemes, i.e., Shade and Remit at the top of ~\autoref{fig:system},
 to counteract \RT.
The detection and mitigation reside within the robot control program.
As a result, they can learn the robot's states and historical records of 
the environment in which the robot is working. 
Using such information (\circled{3} in ~\autoref{fig:system}), the detection
module would report whether the sensor values are compromised or not (\circled{4} in
~\autoref{fig:system}). Upon an alert of detected attacks, the robot control program
cannot rely on the sensor values to proceed moving. Instead,
the mitigation module would be activated to
leverage historical records (\circled{5} in ~\autoref{fig:system}) of obstacles in the environment so as to navigate the robot to complete planned tasks (\circled{6} in ~\autoref{fig:system}).

\section{\RT for Autonomous Service Robot}~\label{sec:fuzzing}
In this section, we first 
model the state transitions of autonomous service robot and
explain the feasibility and procedure of \RT through state composition 
(cf. Section~\ref{sec:composition}). 
Then   
we model \RT, a systematic scheme that effectively 
damages autonomous service robot by fuzzing sensor values
(cf. Section~\ref{sec:robofuzz}).

\subsection{State Compositions of \RT}~\label{sec:composition}

\begin{figure}[t]	
	\centering
	\scalebox{1.00}{\includegraphics[width=\columnwidth]{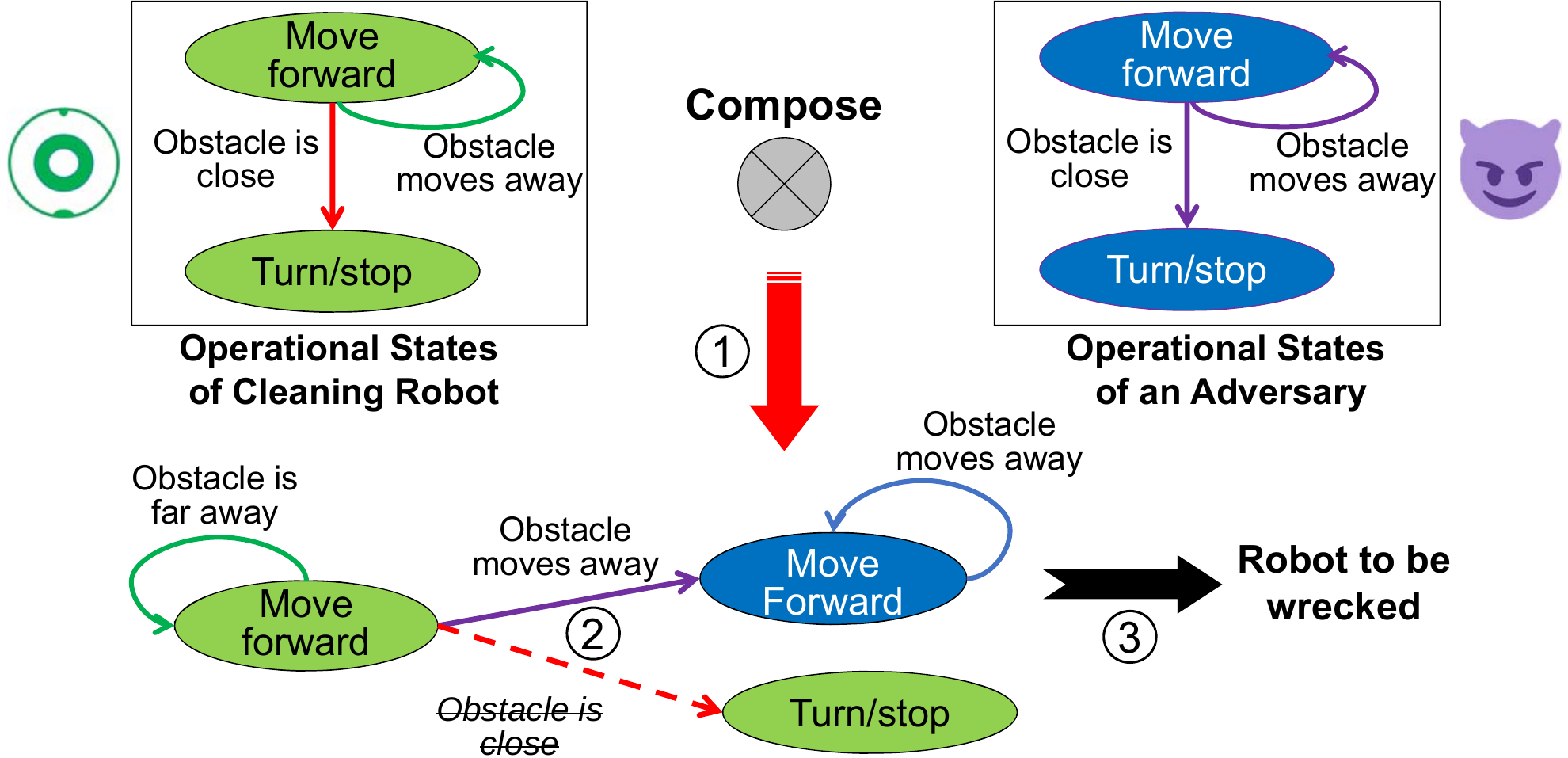}} 
	\caption{An Illustration of State Composition of \RT}~\label{fig:composition}
\end{figure}

An autonomous service robot can be  
 modeled as a finite state machine (FSM). The upper-left
part of Figure~\ref{fig:composition} captures a segment of a simplified FSM
for a cleaning robot. This segment applies to all four
scenarios mentioned in Section~\ref{sec:mot} as it
shows how the cleaning robot proceeds on meeting 
an obstacle that can be either fixed or movable.
Meanwhile, as developers of the robot,
we maintain the FSM (cf. Figure~\ref{fig:composition}) and 
continuously observe the environmental parameters from time to time.
The outcome of \RT hence can be viewed as a {\em composition} of two FSMs 
(\circled{1} in Figure~\ref{fig:composition}). In particular,
once \RT notices a significant change of an 
environmental parameter that is to incur a state transition, 
like the distance to an obstacle decreasing to be very small,
\RT will fabricate a series of rational sensor values
and feed them to the robot control program to make an illusion (\circled{2}
 in Figure~\ref{fig:composition}), e.g., the obstacle moving away.
 By doing so, \RT misleads the robot into
the FSM intended by \RT, which, however, the robot control program will not 
be aware of. Eventually the robot is supposed to be 
wrecked because of hitting the obstacle (\circled{3} in Figure~\ref{fig:composition}).

We note that the main purpose of \RT
is to unveil the vulnerability of robot control program
and in turn compromise the robot
through fuzzing sensor values.  
\RT is an automated procedure.  
It keeps monitoring the states of robot and environmental parameters. At a proper occasion, it activates 
the state composition with faked but rational sensor values  
to deceive the robot control program.

\subsection{Fuzzing Autonomous Service Robots with \RT}~\label{sec:robofuzz}

How \RT compromises an autonomous service robot is modeled as follows.
Because \RT works in line with 
the state of an autonomous service robot and the environment,
it falls into the category of {\em directed fuzzing}. 
Directed fuzzing starts off with a given target, such as damaging the robot or
reducing the robot's work efficacy.
Let these targets form a set,
\setequationmargin
\begin{equation*}
\boldsymbol {T} = \{\tau_0, \tau_1, ..., \tau_i, ..., \tau_{n-1}\},
\end{equation*}
where $\tau_i$ ($0\leq i < n$) is one independent target, e.g., 
to damage the robot, and the value of $n$ depends on the intention of adversaries.
Before fuzzing, we, at the standpoint of adversaries, 
assume that 
the physical states of the robot monitored at runtime form a set, i.e.,
\setequationmargin
\begin{equation*}
\boldsymbol {Z} = \{\zeta_0, \zeta_1, ..., \zeta_k, ..., \zeta_{p - 1}\}.
\end{equation*}
We also assume a thorough understanding of the robot, particularly
all the components embodied 
in the robot, say,
\setequationmargin
\begin{equation*}
\boldsymbol {C} = \{s_0, s_1, ..., s_{l-1}, a_0, a_1, ..., a_{m - 1}\},
\end{equation*}
in which there exist all $l$ sensors 
and $m$ actuators. 
\RT relies on the $l$ sensors to spot the environment.
In addition,
\RT can also utilize actuators
for a target
although we use sensors for illustration
in preceding sections, e.g., by driving wheels faster than usual
towards an obstacle.

To attain a specific target,
\RT must formulate 1) what states and environmental parameters should be 
monitored, 
2) which sensors and actuators in $\boldsymbol {C}$ are useful 
for the target,  
and 3) when (i.e., the aforementioned `appropriate' occasion) 
and how to alter sensor values or actuator commands 
for a detrimental state transition (e.g., transiting
between different curves shown in Figure~\ref{fig:states}).

Hence, 
for a target $\tau_i$, we need  
1) a subset of $\boldsymbol {Z}$, i.e., $\boldsymbol {Z}_i$, which subsumes
states that are useful for $\tau_i$,
2) a subset of $\boldsymbol {C}$, say, $\boldsymbol {C}_i$, which is a list
of essential sensors and actuators for $\tau_i$, and
3) a set $\boldsymbol {V}_i$ in which each element 
includes a tuple for the $j$-th ($0 \leq j < |\boldsymbol {C}_i|$) 
sensor or actuator in $\boldsymbol {C}_i$, i.e.,
\setequationmargin
\begin{equation*}
\left \langle v^{(i)}_{j}, \gamma^{(i)}_{j}, f^{(i)}_{j} \right \rangle.
\end{equation*}
$v^{(i)}_{j}$ is a normal sensor value/actuator command
while $\gamma^{(i)}_{j}$
is a fuzzed sensor value/actuator command. For instance, 
$v^{(i)}_{j}$ and $\gamma^{(i)}_{j}$ fall into the range of $[2, 400]$ (cm)
for an HC-SR04 ultrasonic distance sensor.
Note that both of them 
can also be
a special value $\varnothing$ which stands for the non-existence 
of sensor value/actuator command. $\varnothing$ is useful when there ought to be 
no sensor value/actuator command or adversaries intentionally drop a sensor 
value/actuator command.
The third element in the tuple, 
i.e., $f^{(i)}_{j}$, is a function,
\setequationmargin
\begin{equation}\label{eq:f}
f^{(i)}_{j}: \boldsymbol {Z}_i \times \mathit{Dom} \left (v^{(i)}_{j} \right ) \rightarrow Dom \left ( \gamma^{(i)}_{j} \right),
\end{equation}
where $\mathit{Dom}(x)$ means the domain of $x$. 
Assuming that the robot is at a state
$\zeta \in \boldsymbol {Z}_i$ (e.g., moving forward) 
and one or multiple environmental parameters
are to change,
like when the distance to obstacles, i.e., $v^{(i)}_{j}$, is going to decrease to be 6cm,
$f^{(i)}_{j}$ alters $v^{(i)}_{j}$ to $\gamma^{(i)}_{j}$,
say, from 6cm to 60cm (i.e., making a fixed obstacle `move').
$f^{(i)}_{j}$ hence converts 
a normal sensor value/actuator command or $\varnothing$ 
to be a still rational but harmful value. 
Also 
it may 
replace a sensor value/actuator command with $\varnothing$ to 
hinder the robot control process from interacting with corresponding sensors/actuators.
$f^{(i)}_{j}$ keeps affecting the robot control process until the 
achievement of target $\tau_i$.

Finally, we capture a successful fuzzing procedure for target $\tau_i$  
as: 
\setequationmargin
\begin{equation}\label{eq:entail}
\begin{aligned}[b]
\boldsymbol G_i \vDash \tau_i,
\end{aligned}
\end{equation}
in which $\boldsymbol G_i$ is defined as
\begin{equation}\label{eq:G}
\begin{split}
\boldsymbol G_i = \bigcup_{\zeta \in \boldsymbol {Z}_i} \left \{\langle v^{(i)}, \gamma^{(i)}, f^{(i)} \rangle\ |\ \langle v^{(i)}, \gamma^{(i)}, f^{(i)} \rangle \in \boldsymbol {V}_i \right. \\ \left.\ \land\ \mathit{Dom} \left (f^{(i)} \right ) = \zeta \times \mathit{Dom} \left (v^{(i)} \right )  \right \}. 
\end{split}
\end{equation}
$\boldsymbol G_i$ means that, for every state $\zeta \in \boldsymbol {Z}_i$, 
\RT discovers all tuples related
to $\zeta$ and calls the respective function $f^{(i)}$ to 
fabricate and/or drop
one or multiple sensor values and/or actuator commands for the success of $\tau_i$.

\begin{algorithm}[t]	
	\caption{The $\boldsymbol G_i$ for a Distance Sensor}\label{algo:Gi}
	\begin{algorithmic}[1]
		\Require The target $\tau_i$ for fuzzing;
		\Ensure $\gamma^{(i)}_{j}$ for the distance sensor $s_i$;
		\While {(the robot is working)}
			\State Get the current state $\zeta_k$, and sensor value $v^{(i)}_{j}$;
			\If {($\tau_i$ is to crash the robot)}
				\If {($v^{(i)}_{j}$ is decreasing)} \Comment{Approaching an obstacle}
					\If {($v^{(i)}_{j}$ gradually decreasing)}				
						\State \Comment{Figure~\ref{fig:wall} $\Rightarrow$ Figure~\ref{fig:away}}
						\State When $v^{(i)}_{j}$ is small enough, e.g., $v^{(i)}_{j} < 20$cm, 
						\indent\indent\indent $v^{(i)}_{j} \xrightarrow{f^{(i)}} \gamma^{(i)}_{j}$
						($\gamma^{(i)}_{j}$ continues to increase);
					\ElsIf {($v^{(i)}_{j}$ decreasing more sharply)}
						\State \Comment{Figure~\ref{fig:face} $\Rightarrow$ Figure~\ref{fig:away}}
						\State $v^{(i)}_{j} \xrightarrow{f^{(i)}} \gamma^{(i)}_{j}$ ($\gamma^{(i)}_{j}$ no longer decreases but 
						\indent\indent\indent gradually increases);				
					\EndIf					
				\ElsIf{($\tau_i$ is to reduce the robot's work efficacy)}			
					\If {($v^{(i)}_{j}$ increases and continue increasing)}
						\State \Comment{Figure~\ref{fig:away} $\Rightarrow$ Figure~\ref{fig:wall}}
						\State $v^{(i)}_{j} \xrightarrow{f^{(i)}} \gamma^{(i)}_{j}$ ($\gamma^{(i)}_{j}$ continues to decrease);									

					\ElsIf {($v^{(i)}_{j}$ suddenly increases but then drops)}
						\State \Comment{Figure~\ref{fig:slide} $\Rightarrow$ Figure~\ref{fig:wall}}
						\State When $v^{(i)}_{j}$ suddenly increase,
							$v^{(i)}_{j} \xrightarrow{f^{(i)}} \gamma^{(i)}_{j}$ 
							\indent\indent\indent($\gamma^{(i)}_{j}$ continues to decrease);
					\EndIf
				\EndIf	
			\EndIf			
		\EndWhile
		\State \textbf{Return} $\gamma^{(i)}_{j}$ to replace $v^{(i)}_{j}$ for $\tau_i$;
	\end{algorithmic}
\end{algorithm}
\setlength{\textfloatsep}{0.3cm}
\setlength{\floatsep}{0.3cm}

\paragraph*{Implementing $G_i$}
The implementation of $\boldsymbol G_i$ is based on the rationale discussed in the preceding section (cf. Section~\ref{sec:mot}). 
Algorithm~\ref{algo:Gi} shows the implementation of $G_i$ for a distance sensor
while
the target $\tau_i$ is either to crash the robot or reduce the robot's
work efficacy.
\RT continuously tracks running states of an autonomous service robot 
and waits for a proper time to fuzz the robot ((Lines 1 to 2 in Algorithm~\ref{algo:Gi})). 
For instance, when the sensor value $v^{(i)}_{j}$ is gradually decreasing (Line 4), \RT realizes that there is a fixed obstacle ahead.
Therefore, to crash the robot ($\tau_i$ at Line 3), 
the $G_i$ function would generate sensor values, i.e., 
$\gamma^{(i)}_{j}$, which continue increasing to resemble a leaving obstacle (Line 7).
By doing so, \RT aims to use faked sensor values to change the scenario shown by Figure~\ref{fig:wall} to the one in Figure~\ref{fig:away}. 
Algorithms~\ref{algo:Gi} also shows how to convert scenarios for other types of obstacles (
Lines 8 to 10, Lines 12 to 15, and Lines 16 to 18).

\section{Attack Detection with Shade}\label{sec:detection}

\RT provides a way to initiate successful attacks to an autonomous service robot.
In this section we will consider
how to efficiently detect attacks.
We first investigate possible attack models which, once integrated with \RT,
would carry the robot into misbehaving states. Accordingly
we look into three classic detection methods, 
and develop a hybrid one with wider coverage, higher accuracy and less overhead.

\subsection{Classic Detection Methods}

\paragraph*{\textbf{Fingerprinting}}
Hardware devices
have their unique physical characteristics~\cite{fingerprint:CIDS:Security-2016}, 
i.e., {\em fingerprints}, such as 
the sensor latency (i.e., response time).
Assuming that attackers 
fabricate and send fake sensor values via Wi-Fi,
the sensor latency observed by the control process should 
be extraordinary as network latencies are generally
one or two orders of magnitude longer than typical sensor latencies.
Take the ultrasonic distance sensor (HC-SR04) for example.
Its sensor latency mostly falls in a range of 2ms to 12ms.
By contrast, the network latency under TCP and UDP protocols varies between 
200ms and 250ms. If the robot control process has learned a sensor's normal latency, 
it is able to detect an attack 
that delivers sensor values through the network.

Fingerprinting is advantageous with its simplicity and low overhead.
But it has limited usages.
Given a sensor 
working in a periodical or proactive mode, 
the robot control process cannot measure 
its sensor latency for validation.

\paragraph*{\textbf{Cross-reference Validation (CRV)}} 
CRV leverages
information from two or more sources to cross-check for verification.
The challenge in using CRV for an autonomous service robot 
is that every sensor might be compromised
and  
using different sensors for cross-checking is unreliable. Also,
not many sensors are installed in a small service robot 
for similar purposes.
CRV must use some information that attackers are unaware of. 
Let us still use the distance sensor for example.
A cleaning robot 
can make historical records of the positions of stationary obstacles 
in a normal working routine.
In fact, some
iRobot Roomba robots draw a map of the space 
they have cleaned~\cite{robot:roomba-map:2018}.
Such historical records 
can be secured and used as the norm to validate 
distance sensor values.
If the distance sensor gives a value that badly
violates historical records,
 CRV can indicate the occurrence of an attack.

Compared to fingerprinting, CRV can
detect attacks that compromise sensors working in the proactive mode
since CRV cross-checks by exploiting extra historical 
records.
However, CRV requires continually
tracking the robot's motion so as to refer to 
the correct record. 
Also,
the accuracy of CRV is not very high 
because records have been approximately made~\cite{robot:localization-detection:RAS-2018}.

\begin{figure}[t]
	\centering
	\scalebox{1.00}{\includegraphics[width=\columnwidth]{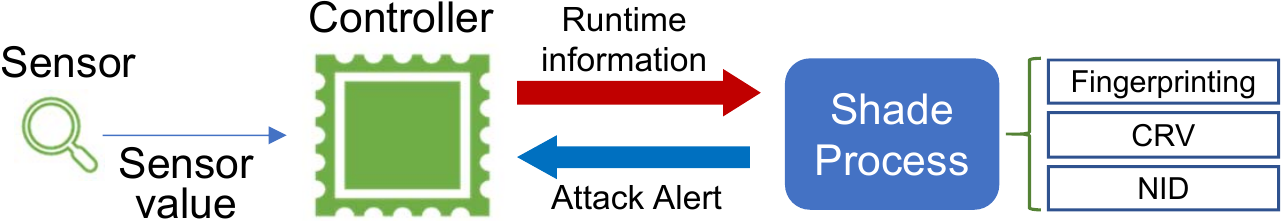}} 
	\caption{An Illustration of Shade with Robot Controller}~\label{fig:shade}
\end{figure}

\paragraph*{\textbf{Network intrusion detection (NID)}}
NID performs an analysis over 
the behavior, payload and contents of inbound and outbound network packets~\cite{security:network:SP-2010}. 
As attackers remotely attack the robot via network,  
NID should be practical.
Given a service robot working in a normal routine,
packets exchanged between it and a legitimate user must
follow a regular pattern and the network 
payload should not change largely. 
But when attackers undertake to obtain and alter sensor values,
they would bring about unusual network packets,  
 either in a large
quantity or with abnormal contents.
An independent process monitoring the network traffic
should detect such breaches.

A major drawback of NID is its high cost in computation
and energy. Therefore, in an autonomous service robot powered 
by a battery, NID should
be periodically called 
for energy-efficiency~\cite{CPS:battery:TDAES-2017}. 
Also, NID cannot capture all attacks although they go through 
the network interface. 
Consider suspension attacks. If  
attackers manage to compromise a sensor just at 
the first try with few packets, NID 
might neglect such an attack.

\subsection{The Design of Shade}

Each of the aforementioned methods has its strengths and limitations.
We have developed a hybrid method called the {\em \underline{sha}dow
\underline{de}tector} (Shade). 
Shade acts as a shadow process of the robot control process and closely communicates with 
the latter to avoid missing any attack imprint.
~\autoref{fig:shade} illustrates how the Shade process collaborates 
with robot control process through inter-process communication.
The control process provides runtime information to Shade,
such as the motion trace, sensor values, sensor latencies, etc.  
On the other side, Shade swiftly informs the control process in case of attacks.

\begin{algorithm}[t]	
	\caption{The Shade Process ({\tt Shade()})}\label{algo:shade}
	\begin{algorithmic}[1]
		\Require  A request for attack detection $p$ with runtime information;\Comment{ $p$ may contain the sensor mode $\mu$, the sensor latency $\lambda$, the current location of robot $\zeta$, etc.}
		\Ensure An attack alert $\gamma$ \Comment{ $\gamma$ will be either {\bf True} or {\bf False} }
		\If {($p$ is with sensor information)}
			\If{($\mu$ is \textit{PASSIVE})} \Comment{Robot actively demands sensor value}
				\State \Comment{ Shade calls fingerprinting method with sensor latency}
				\State{$\gamma :=$ {\tt fingerprinting\_check($\lambda$)}};
			\Else \Comment{ The sensor reports to robot periodically or in emergency}
				\State \Comment{ Shade calls CRV first with robot location, then NID}
				\State{$\gamma :=$ {\tt CRV\_check($\zeta$)} {\bf Or} {\tt NID\_check()}};
			\EndIf
		\Else \Comment{ Robot controller queries without sensor information}
				\State \Comment{ Shade calls NID method}
				\State{$\gamma :=$ {\tt NID\_check()}};			
		\EndIf			
		\State \textbf{Return} $\gamma$ to the robot controller process;
	\end{algorithmic}
\end{algorithm}
\setlength{\textfloatsep}{0.3cm}
\setlength{\floatsep}{0.3cm}

Shade is a hybrid mechanism of fingerprinting, CRV and NID so as to
achieve wide coverage, high accuracy and low overhead. 
Algorithm~\ref{algo:shade} describes the main procedure of Shade.
The robot control process sends a request for attack detection either
in an on-demand or periodical way and the Shade process
returns whether an attack is happening or not.
If Shade receives a request with sensor information 
(Lines~1 to 8 in Algorithm~\ref{algo:shade}), 
it first determines the working mode
of the sensor. Given a sensor working at a passive mode 
with a measurable latency,
Shade prefers the fingerprinting method that 
comes with low cost but high accuracy (Lines 2 to 4).
However, as to a sensor working
in a proactive or periodical mode, 
Shade calls CRV to validate the sensor value  
against historical records (Lines 5 to 8); nevertheless,
due to the accuracy of CRV, Shade may use NID
for double check with a short-circuiting logical \textbf{Or} operator (Line 7).
Moreover, the robot control process may ask Shade without any sensor
information. For example, the control process can consult Shade every five
seconds. In this case, Shade needs to execute NID that finds out abnormal
network traffics (Lines 9 to 11).
In the end, Shade timely notifies the robot control process with a detection
result (Line 13).

Shade can detect various attacks and it 
is beyond just integrating three methods in one process. First, Shade 
explores the {\em context} provided by the robot control process for attack detection. 
Generic NID can also detect the most attacks 
but with high cost for self-learning and frequent computations. Shade, however, 
gains legitimate network behaviors shared by the robot control process, 
which surely entails higher accuracy and less overhead. Second, Shade considers 
the pros and cons of three methods and complement them for wider coverage. Like
at Line~7 of Algorithm~\ref{algo:shade}, Shade makes NID recheck
if CRV generates a false result because of the latter's accuracy.

\begin{algorithm}[t]	
	\caption{The Mitigation Process ({\tt Remit()})}\label{algo:mitigation}
	\begin{algorithmic}[1]
		\Require A switch from normal mode to safe mode for mitigation.
		\Ensure A completion of task, or a switch back to normal mode.
		\Repeat 
		\State Change/keep the robot moving at a lower speed;			
		\State \Comment{ Navigate the robot with historical records used by {\tt CRV\_check()}}
		\State Call {\tt Navigate\_with\_historical\_records()};
		\State Try to reset corresponding sensor;
		\State Call {\tt Network\_block\_attacks()} to block attackers;
		\If {(attackers are successfully blocked)} 
		\State {\bf Return} back to normal mode;
		\EndIf
		\State \Comment{ A dynamic obstacle (e.g., a pet) might appear}
		\If {(Robot cannot move with no obstacle recorded)} 
		\State Play sound to drive the person/pet, and wait 1 second;										
		\EndIf
		\State Continue moving with historical records;		
		\If {(the scheduled task is completed)}
		\State {\bf Return} a completion to the robot control process;
		\EndIf
		\Until{({\tt Shade()} returns {\bf False})}; \Comment{ No attack any longer}
		\State {Switch back to the normal mode of robot control process};
	\end{algorithmic}
\end{algorithm}

\section{Mitigation with Remit}\label{sec:migitation}

Once Shade detects  
any attack affecting an autonomous service robot,
we must mitigate the attack's impact. A straightforward solution
is to halt the robot immediately. 
However, a shutdown of the robot badly loses its work efficacy 
because the robot is supposed to have a scheduled task, like tidying a room.
As a result, we need a mitigation algorithm that retains as much work 
efficacy as possible for the robot being attacked. In particular,
the mitigation algorithm ought to take into account two issues.
First, an autonomous service robot significantly differs from
stationary CPS and handheld smartphones
as the robot needs to move itself to work. 
Since the distance sensor is not reliable due to attacks, 
how to navigate the robot to continue its motion 
must be resolved. Second, because of the limited resources of a small service robot, 
including the computation capability and energy supply, 
the mitigation algorithm should be lightweight and cost-efficient.

Regarding the two challenges, we have 
designed a mitigation algorithm, namely 
{\em \underline{re}taining-oriented 
	\underline{mit}igation} ({\em Remit}), to achieve
the least loss of work efficacy for an autonomous service robot. 
One noteworthy point of Remit is that, it reuses the 
historical records used by Shade in detecting attacks with CRV, which
not only preserves the motion of robot, but also avoids any extra cost
for enabling the navigation.
Algorithm~\ref{algo:mitigation}
captures the procedure of Remit. 
We define the robot without being attacked 
is in the {\em normal} mode. Remit switches the robot to {\em mitigation} mode once Shade
detects an attack. On entering the safe mode, the robot first decelerates its speed 
(Line 2 of Algorithm~\ref{algo:mitigation}). This helps it have more time to 
respond to an emergent object, say, an obstacle.
Then,  
Remit leverages the historical records to navigate the robot (Lines 3 to 4). 
Since these records are not very accurate, Remit  
tries to repair the compromised sensor through resetting (Line~5) and
calls the network module to 
block attackers (Line~6).
If the attackers are successfully blocked,  
Remit will switch the robot back to the
normal mode (Lines 7 to 8). Remit also needs to deal with a dynamic
obstacle (e.g., a pet or person) 
if the robot cannot move at a time 
but no obstacle was recorded (Lines 10 to 13). Remit
alerts the pet or person by playing a sound (Line 12) and continues moving (Line 14). If the robot 
completes the scheduled task, Remit stops the robot 
(Lines 15 to 17). Otherwise, Remit  
repeats the aforementioned steps until Shade detects no 
attack any more (Line 18).

Remit attempts to guarantee the work efficacy of the robot.
Since the robot needs to move at a lower velocity, 
the time needed to complete a planned task might become longer. 
However, with regard to the robot being under attack, 
such additional time cost is insignificant and acceptable.

\section{Evaluation}~\label{sec:evaluation}
In this section, we would evaluate \RT, Shade, and Remit to 
answer following questions.
\begin{enumerate}[label={\arabic*})]
	\item Does \RT manage to compromise an autonomous service robot? Compared to other fuzzing approaches,
	does \RT embrace a higher success rate?
	\item Is Shade able to detect most of the attacks initiated through \RT?
	\item Can Remit retain the work efficacy of service robot when mitigating the attacks detected by Shade? 
\end{enumerate}
In brief,
we first present experimental setup and evaluation results regarding
the competence of \RT in compromising a real-world
cleaning robot with two attack targets. 
Then we test Shade and Remit to show their effectiveness in detecting two attack models 
and retaining the work efficacy of robot.

\subsection{Evaluation Setup}
\paragraph*{Platform}
We use the aforementioned iRobot Create 2~\cite{robot:irobot-create2:2015}
as the platform for evaluation.
We have prepared a control program in Python 3 that runs in the Raspberry Pi 3 Model B+.
The default velocity of the robot is set to be 50mm/s.
The path planning of the robot follows the classic zigzag fashion.
The main sensor used for the path planning is 
an ultrasonic distance sensor (HC-SR04) installed 
in front of the robot. The sensor
can detect an obstacle from 2cm to 400cm.
As mentioned in Section~\ref{sec:bg},
in the robot control program, we configure
the sensor to work in different modes to suit different attack models.
In the passive sensor mode, the control program asks for the sensor value. 
In the proactive sensor mode, the sensor warns the control process 
if an obstacle is nearby or periodically.

\begin{figure}[t]				
	\centering
	\includegraphics[width=\columnwidth]{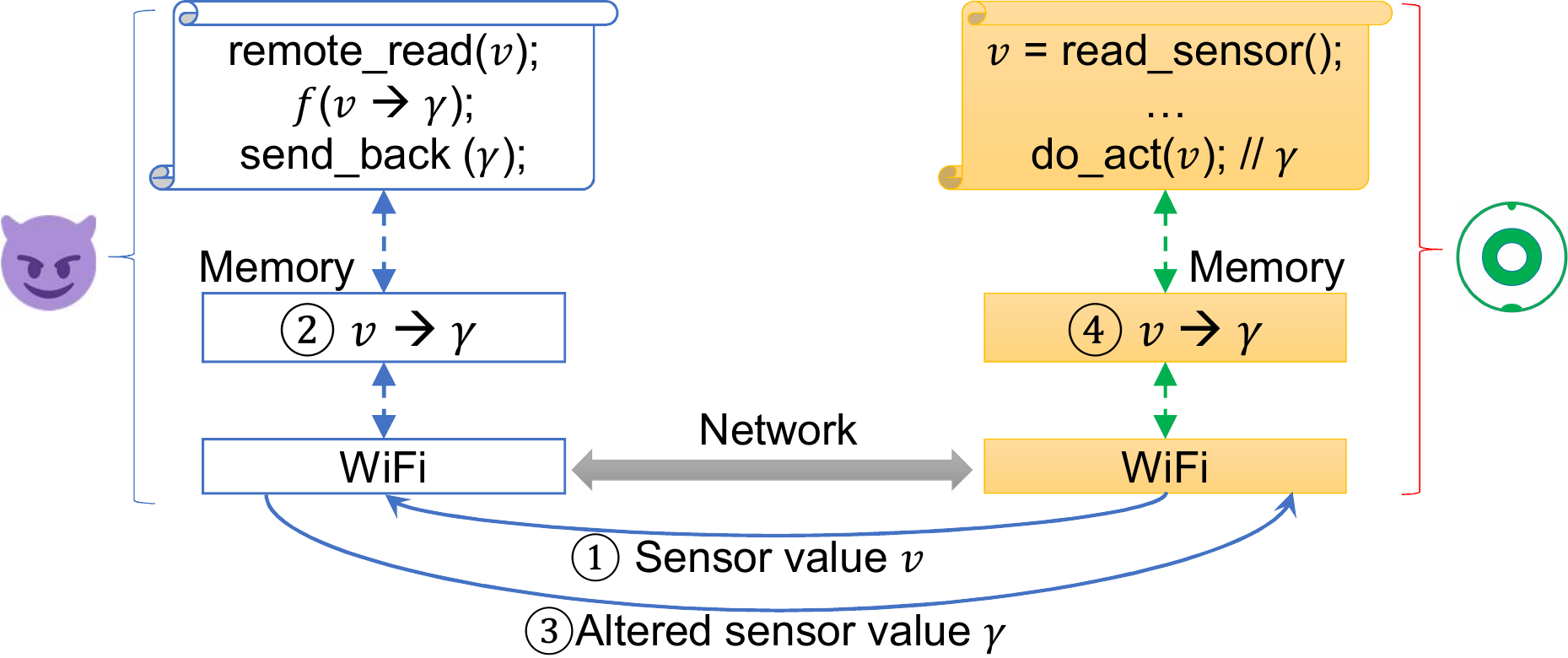}  
	\caption{An Illustration of Initiating an Attack}~\label{fig:impl}	
\end{figure}

As to the attacker side,.
we have implemented \RT with two attack models (cf. Section~\ref{sec:model}). 
In light of the analyses in Section~\ref{sec:model},
we set the sensor mode to be passive
for fabrication attack model.
For the suspension attack model, we choose 
the proactive mode.

\paragraph*{Implementation}
In order to manipulate 
the \iRobot Create 2, we make attack programs in
a computer  
with Ubuntu 18.04. 
We first exploit the attack vector of WiFi interface of Raspberry Pi
as the attack surface to invade the robot.
Today many users still use default or simple
passwords  
or their credentials are stored
in plain text~\cite{robot:industrial-security:SP-2017,robot:Mi-hack:2018}.
For a Raspberry Pi with Raspbian, its default username/password
are `pi/raspberry'. After we successfully access the robot,
we start to compromise it.
Figure~\ref{fig:impl} exemplifies the process of altering one
sensor value. As adversaries, we fetch the sensor value $v$
through network (\circled{1} in Figure~\ref{fig:impl}), and
then alter it to be $\gamma$ via the function $f$ 
(\circled{2} in Figure~\ref{fig:impl}).
After sending back $\gamma$ and replacing $v$ (\circled{3} and 
\circled{4} in Figure~\ref{fig:impl}), the robot control program would 
proceed with $\gamma$ instead of $v$.
At runtime, \RT frequently reads $v$, but only when it
perceives an appropriate opportunity, like the robot approaching
a wall, will it call $f$ and send faked $\gamma$ back to mislead
the robot control program.

\begin{figure}[t] 		
	\centering
	\subfigure[The Use Case Scenario for Testing \RT]{\includegraphics[width=0.9\columnwidth]{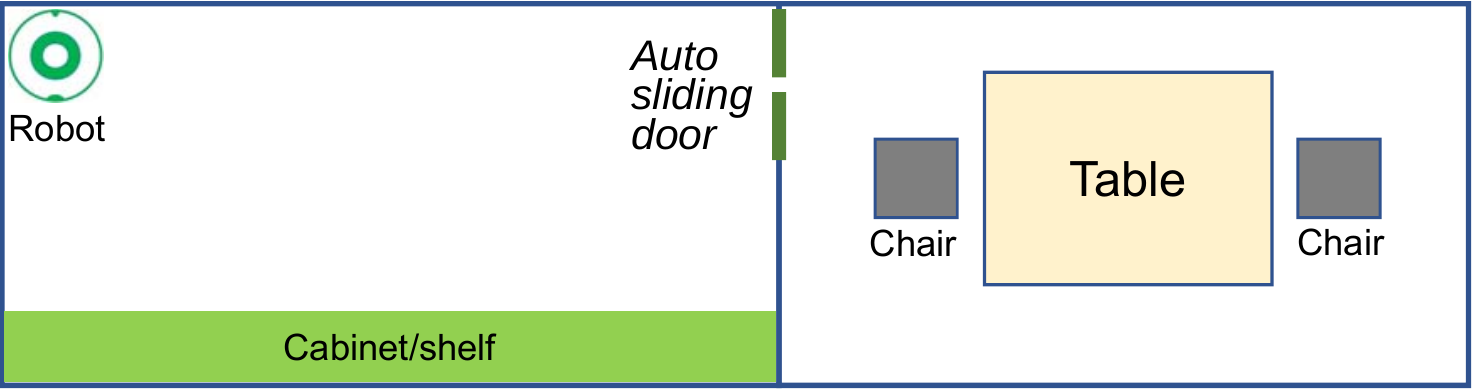}~\label{fig:room}}
	\vfill
	\centering
	\subfigure[The Use Case Scenario for Testing Shade and Remit]{\scalebox{1}{\includegraphics[width=1\columnwidth]{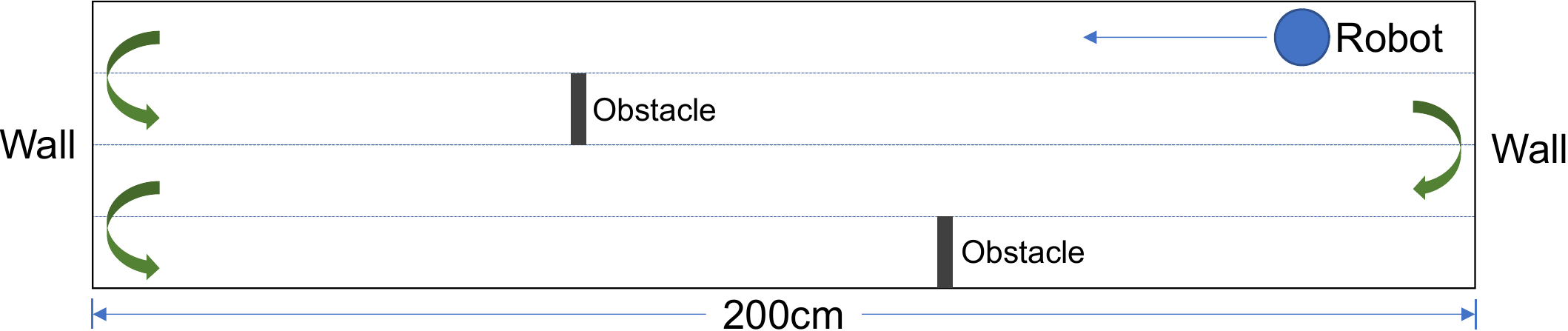}}~\label{fig:case}}			
	\caption{An Illustration of Use Case Scenarios for Evaluation}
	\label{fig:scenarios}	
\end{figure}

\paragraph*{Configuration}
We have made two scenarios to test \RT, Shade and Remit, respectively.
Figure~\ref{fig:room} shows the scenario we have used for testing \RT. It 
has two rooms that are connected by an automatic sliding door. 
The cleaning robot needs to clean both rooms
starting from the top-left corner. When 
the robot is working, we  
try to compromise it using three fuzzing methods with two attack targets: 
1) to damage the robot by crashing it
to a hard obstacle (e.g., wall or cabinet), and 2) to reduce the robot's
work efficacy by preventing it from entering and cleaning the right room.
As to three fuzzing methods, the first one is Radamsa fuzzing sensor values
for the control program,
the second one is random fuzzing that
initiates an attack at a random time with a hazardous
 alteration of sensor values (e.g.,
changing $v$ of 10cm to $\gamma$ of 60cm), and the third one is \RT.
For both attack targets, we conducted 30 trials for a fuzzing method.
We define the success rate as the fraction of the number of successful attacks to
30 trials in percentage. 
We note that besides Shade,
the robot control program can  
rule out anomalous distance sensor values, e.g., ones that fluctuate greatly,
and subsequently reset the sensor.
 
Figure~\ref{fig:case} captures the scenario we would use to test Shade and Remit.
The reason why we evaluate them in a scenario different from Figure~\ref{fig:room}
is that we need a quantitative presentation to measure the work efficacy of robot
in case of attacks. As mentioned,
we use the quantitative success rate to show the effectiveness of \RT.
On the other hand,
the width of the room in Figure~\ref{fig:case} is 200cm that falls into the range of 
HC-SR04 ($\leq$400cm);
the cleaning robot would cruise in the room, so we can record the exact distance
a robot cleans with and without attacks.
To avoid physically crashing the robot
into walls or obstacles due to attacks,
we set a safe distance to be 20cm. 
In other words,
without any attack,
when the distance to a wall or obstacle drops below 20cm, 
the robot should stop moving and turn left or right; however,
on a successful attack, the robot spins itself in front of 
an obstacle to indicate that it is being attacked 
instead of really colliding with the obstacle.
Concerning the safe distance and the diameter of robot, 
the robot would clean an estimate of 570cm overall in the room of Figure~\ref{fig:case}.
In addition, for the use of fingerprinting and CRV, we 
have run the robot without any attack to collect
sensor latencies and historical records of obstacles.

\clearpage
\subsection{Target 1 for \RT: Damaging the Robot}~\label{sec:scenario1}

In order to damage the robot, fuzzing methods must let the robot crash into a fixed 
obstacle in Figure~\ref{fig:room}.
Note that the robot was not really damaged in trials 
but would play a special sound 
to indicate
it was enforced to be 
within 5cm to an obstacle.
~\autoref{tab:t1} shows the number of successful attacks out of overall 30 trials for 
three fuzzing methods when they tried to achieve the target of damaging the robot.
Radamsa failed in all trials
because the robot control process certainly 
refused sensor values fuzzed by it 
as they evidently deviate from normal sensor values expected 
in the environment shown in Figure~\ref{fig:room}.
As to random fuzzing, with regards to multiple stationary walls and furnitures 
in Figure~\ref{fig:room}, if it launched an attack at a moment when, though being randomly picked,
the robot was approaching closely to any wall or furniture,
the fuzzed sensor values might make the robot hit the obstacle and in turn
attain the attack target. Whereas, since random fuzzing acts based on randomization,
 the success rate is low as confirmed by the experimental results (5 out of 30 trials).

\begin{table}[b]
	\centering
	\caption{The number of successful trials and success rates of three fuzzing methods to achieve the 1st target}\label{tab:t1}
	\resizebox{\linewidth}{!}{
		\begin{tabular} {|c|c|c|c|}
			\hline			
			\rule{0pt}{5pt}  Fuzzing Method &  Radamsa &  Random fuzzing &  \RT \\
			\hline\hline
			The number of  successful trials & \multirow{1}{*}{ 0} & \multirow{1}{*}{ 5} & \multirow{1}{*}{ 30}  \\ \hline
			\multirow{1}{*}{ Success rate} & \multirow{1}{*}{ 0\%} & \multirow{1}{*}{ 16.7\%} &
			\multirow{1}{*}{ 100\%}   \\ \hline
		\end{tabular}
	}
\end{table}
\normalsize

\begin{figure}[t]
	\centering
	\subfigure[Robot moving towards a wall without any attack]{\includegraphics[width=0.475\columnwidth]{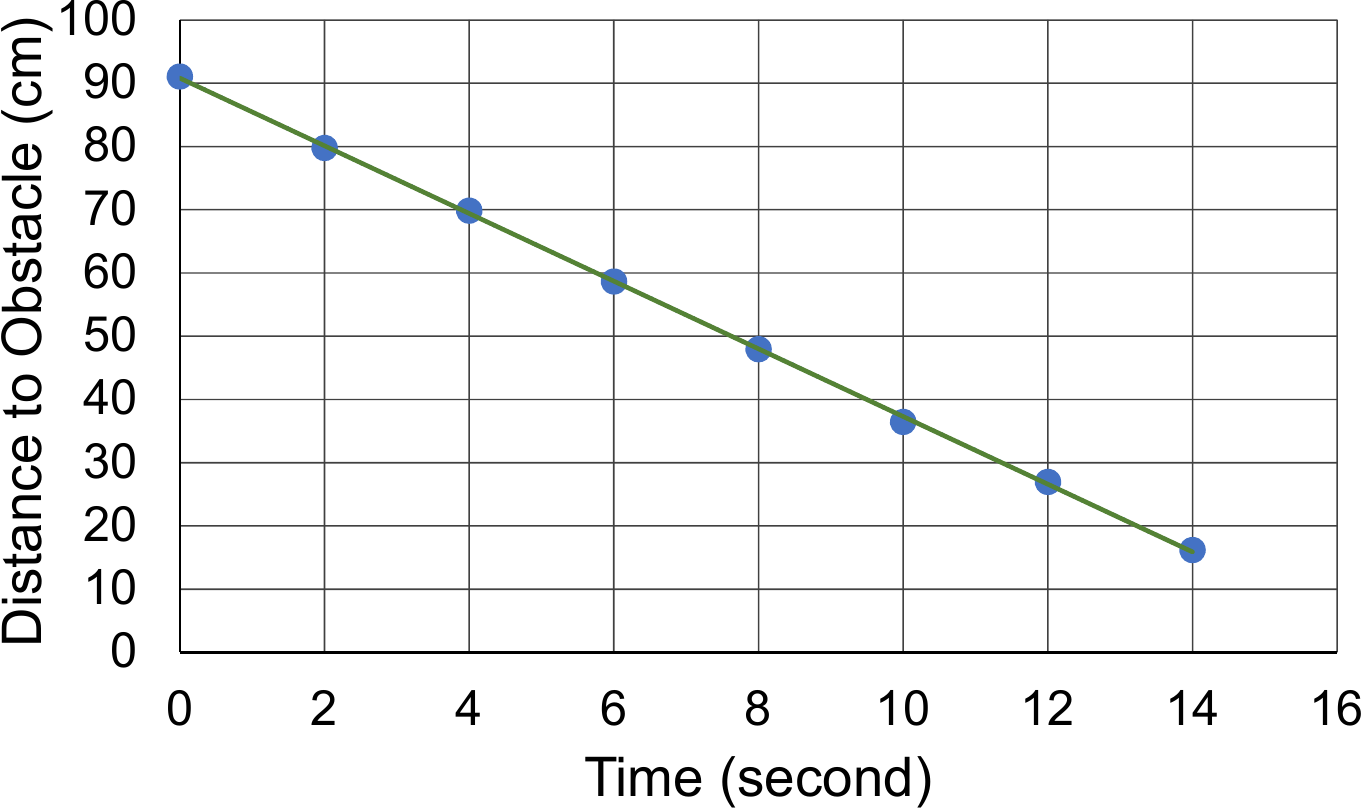}~\label{fig:nwall}}
	\hfill
	\subfigure[Robot deceived with the appearance of dynamic obstacle]{\includegraphics[width=0.475\columnwidth]{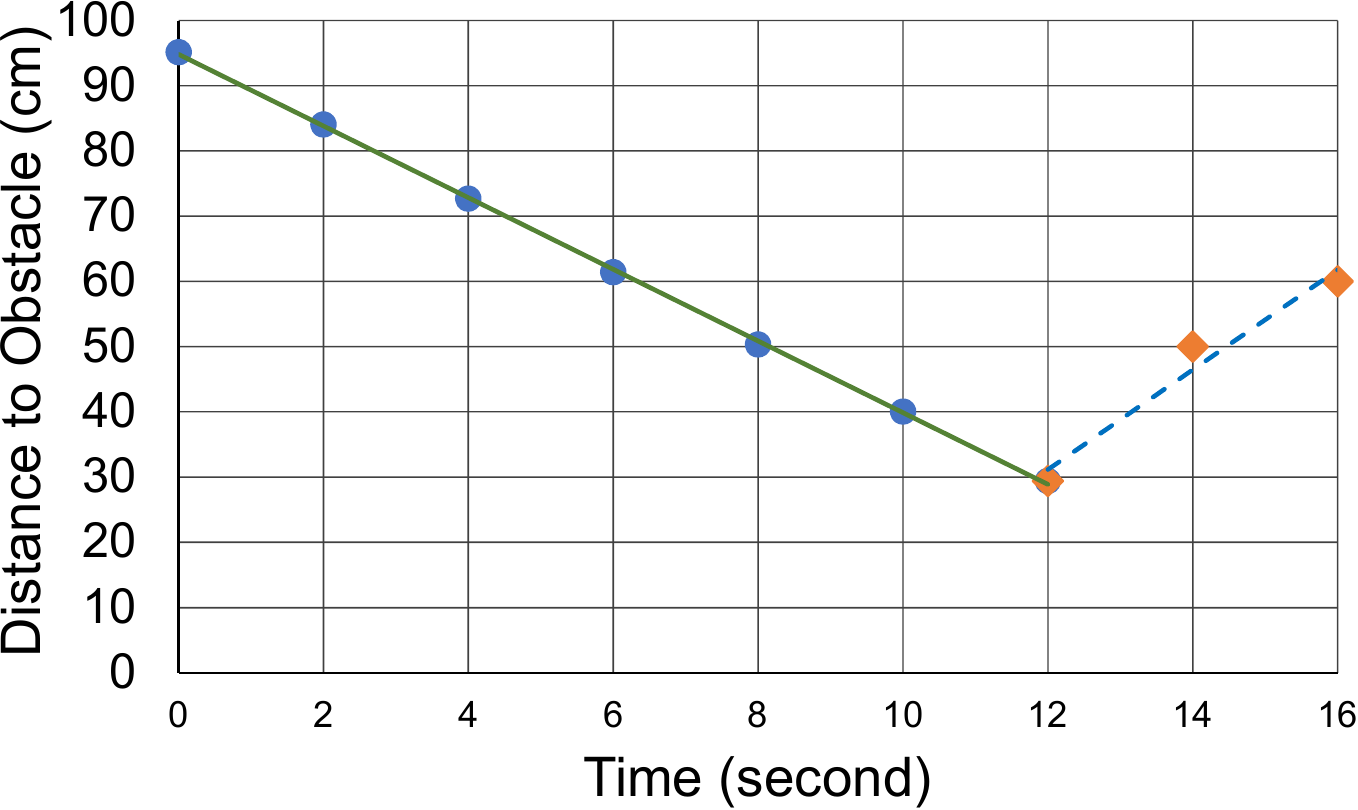}~\label{fig:fwall}}
	
	\caption{A Comparison between Distance Sensor Values with and without \RT when damaging the robot}\label{fig:rt-wall}
\end{figure}

\begin{figure}[t]		
	\centering
	\subfigure[Robot moving towards automatic sliding door]{\includegraphics[width=0.475\columnwidth]{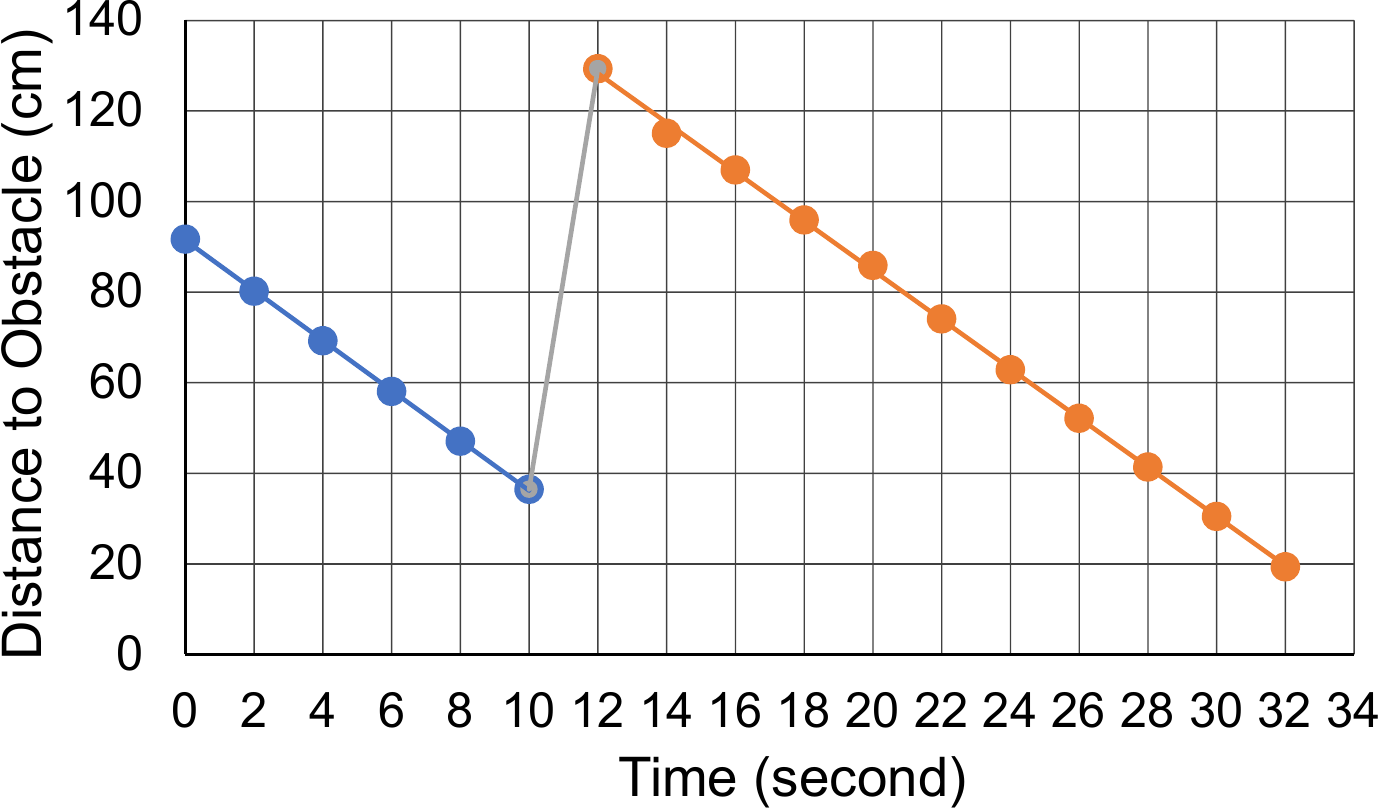}~\label{fig:ndoor}}
	\hfill
	\subfigure[Robot stopped with auto sliding door changed to be an immovable wall]{\includegraphics[width=0.475\columnwidth]{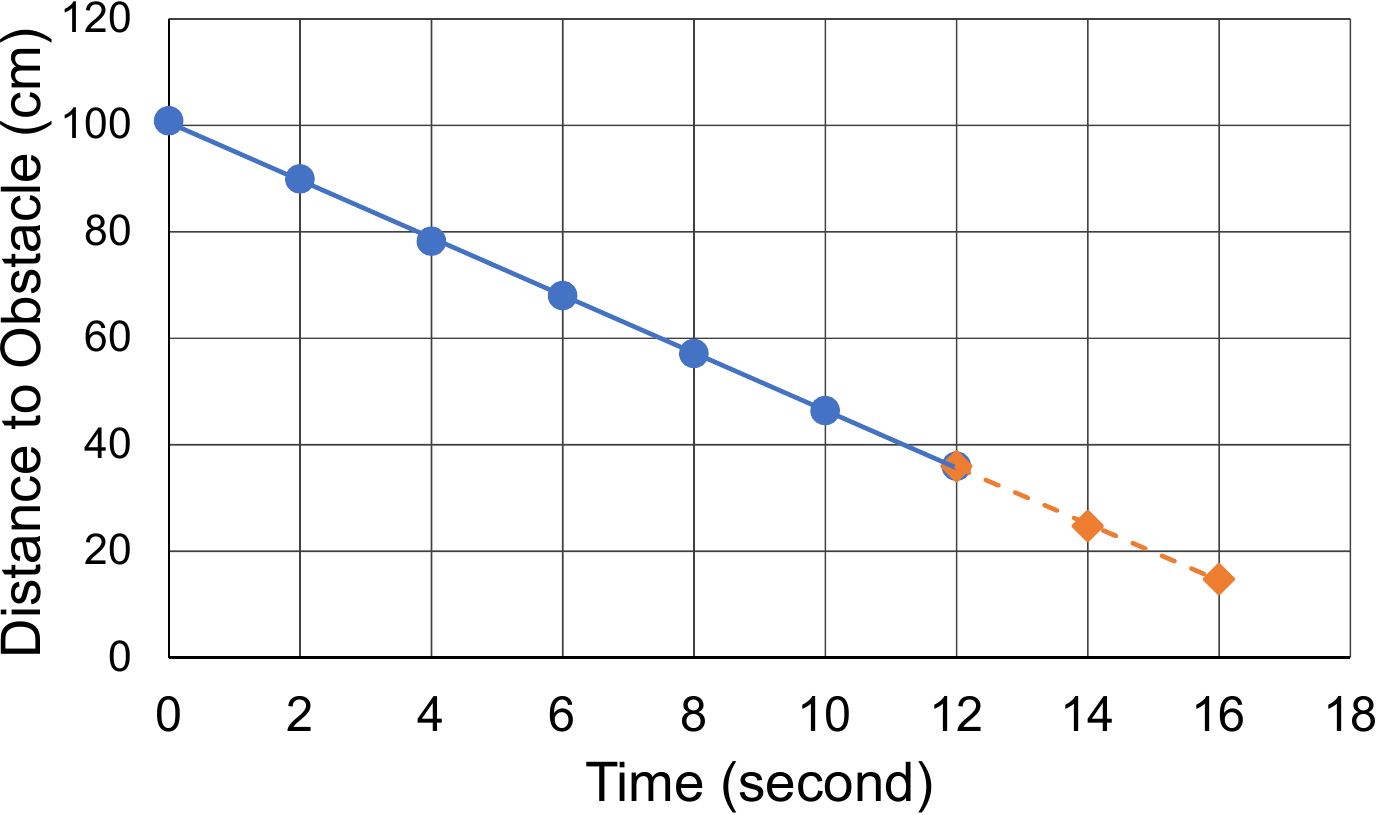}~\label{fig:fdoor}}	
	\caption{A Comparison between Distance Sensor Values with and without \RT when prematurely stopping the robot}\label{fig:rt-door}
\end{figure}

On the other hand,
\RT successfully damaged the robot in all 30 trials. Because
\RT continued to observe the environment and monitor the state of robot,
at a proper occasion, it would generate sensor values that 
brought the robot from the curve in Figure~\ref{fig:wall} to 
the one in Figure~\ref{fig:away}. 
For a thorough comparison,
we have collected distance sensor values in a normal 
routine without any attack and 
when \RT took effect in one trial.
Figure~\ref{fig:nwall} indicates that the sensor values from the normal routine
well fit in a decreasing linear curve.
On the other hand,
in Figure~\ref{fig:fwall}, the solid linear curve links genuine sensor values 
before the attack initiated by \RT 
and the dashed line fits sensor values that impaired the robot.
The two diagrams in Figure~\ref{fig:rt-wall} 
clearly verify the capability of \RT.

\subsection{Target 2 for \RT: Reducing the Work Efficacy of Robot}~\label{sec:scenario2}

To reduce the work efficacy of the cleaning robot,
we called three fuzzing methods to  
hinder the robot from tidying the right room.
In other words, 
after the robot finished cleaning up the left room, the robot 
should not cross the automatic sliding door due to attacks. 
~\autoref{tab:t2} shows that \RT achieves a success 
rate of 93.3\% while the rates for Radamsa and random fuzzing are still low.
Note that the success rates for both random fuzzing and \RT drop compared to that with 
the first target. The reason is, on damaging the robot, 
both fuzzing methods could find a number of static obstacles to leverage,
but there is only one automatic sliding door connecting two rooms.  
Even so, \RT managed to sense the existence of automatic sliding door, and
successfully changed sensor values in the most trials (28 out of 30) 
to be decreasing ones that emulated the door as an immovable wall.

We again tracked sensor values when the robot was going through
the sliding door without attack (cf. Figure~\ref{fig:ndoor}). Also in one successful
trial, we recorded sensor values the 
control process received 
before and after \RT launched the attack (cf. Figure~\ref{fig:fdoor}).
As observed in Figure~\ref{fig:rt-door}, after 12s,
\RT effectively deceived the robot which subsequently
stopped in front of the automatic sliding door.

\begin{table}[b]
	\centering
	\caption{The number of successful trials and success rates of three fuzzing methods to achieve the 2nd target}\label{tab:t2}
	\resizebox{\linewidth}{!}{
		\begin{tabular} {|c|c|c|c|}
			\hline
			\rule{0pt}{5pt}  Fuzzing Method &  Radamsa &  Random fuzzing &  \RT \\
			\hline\hline
			The number of successful trials & \multirow{1}{*}{ 0} & \multirow{1}{*}{3} & \multirow{1}{*}{ 28}  \\ \hline
			\multirow{1}{*}{ Success rate} & \multirow{1}{*}{ 0\%} & \multirow{1}{*}{ 10.0\%} & \multirow{1}{*}{ 93.3\%}   \\  \hline
		\end{tabular}
	}
\end{table}
\normalsize

\begin{table*}[t]
	\centering
	\caption{Detection Results of Four Detection Methods under Two Attack Models\label{tab:detect}}
		\resizebox{\linewidth}{!}{
	\begin{tabular} {|l|c|c|c|c|c|c|r|c|}
		\hline
		\multirow{2}{*}{Attack Model} & \multicolumn{4}{c|}{Number of Trials Detected} & \multicolumn{4}{c|}{Average Reaction Time for Detection (unit: second)}  \\ \cline{2-9}
		& Fingerprinting & CRV & NID & Shade & Fingerprinting & CRV & NID & \multicolumn{1}{c|}{Shade} \\ \hline\hline
		Suspension Attack Model & 0 & 10 & 0 & 10 & Nil & 27.1 & Nil
		& 27.2 \\ \hline			
		Fabrication Attack Model & 10 & 10 & 10 & 10 & 0.6 & 0.8 & 10.1 & 0.6 \\ \hline
	\end{tabular}
		}	
\end{table*}
\normalsize

\begin{figure}[t]
	\centering
	\subfigure[Suspension Attack Model]{\includegraphics[width=0.48\columnwidth]{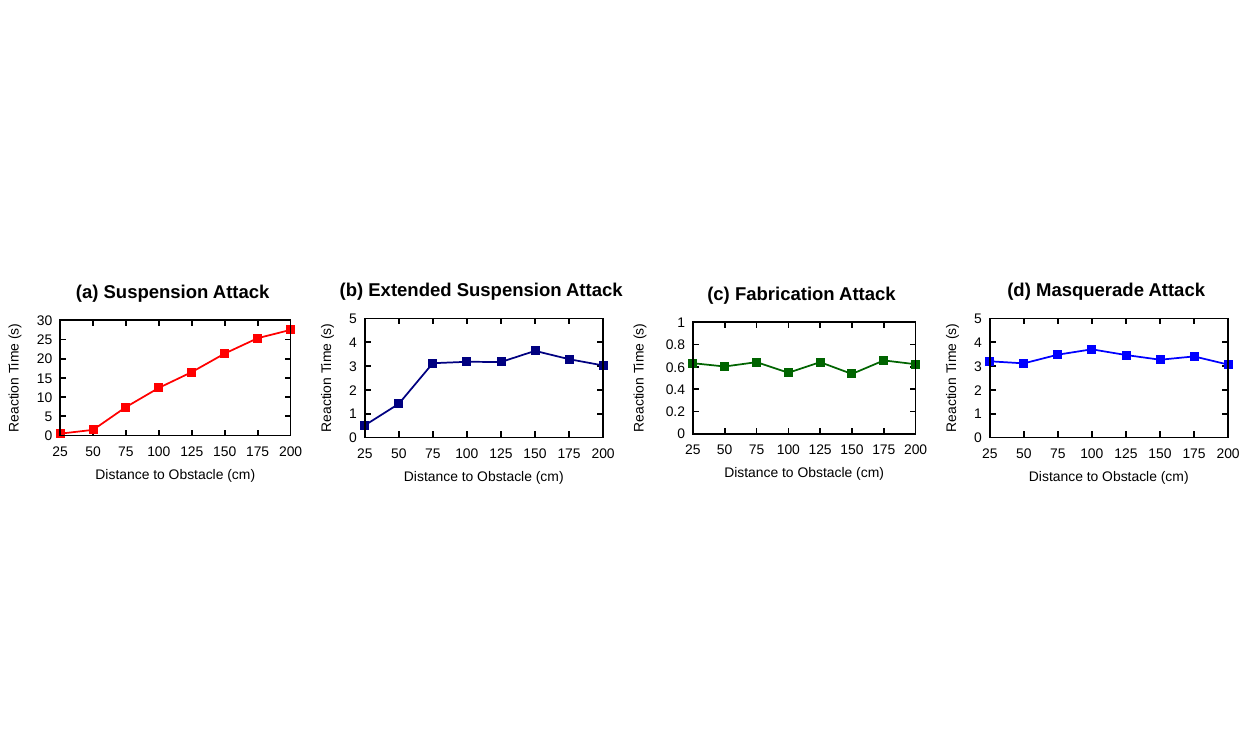}~\label{fig:s-range}}
	\subfigure[Fabrication Attack Model]{\includegraphics[width=0.48\columnwidth]{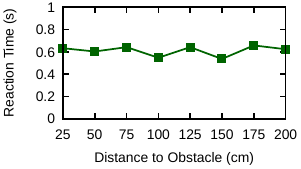}~\label{fig:f-range}}
	\caption{The Reaction Time of Shade to Attacks at Different Distances to Obstacle with Two Attack Models}\label{fig:range}
\end{figure}

\subsection{Detection Results of Shade}

We compared Shade to fingerprinting, CRV and NID methods. 
We used \RT to initiate attacks in line with 
two aforementioned attack models, i.e., suspension and fabrication attacks.
For each attack model,
a detection method
underwent ten trials of attacks. So in all we performed $2 \times 4 \times 10 = 80$ trials
regarding the composition of detection methods and attack models.
Every trial was triggered at the startup of the robot, which means 
the robot is at the top-right corner as shown in Figure~\autoref{fig:case}.
We did so because an attack at the very beginning may
incur the most challenges for a detection method, especially when the 
sensor works at a proactive mode reporting boolean values.
We use two metrics to evaluate the effect of detection. 
One is the number of trials that a detection method successfully detected 
under an attack model. The other one is the average reaction time of ten trials
for a detection method under each attack model.

~\autoref{tab:detect} summarizes the results collected in 80 trials.
Shade has successfully detected all trials 
while
the limitations of other three methods are evident. For example,
fingerprinting is competent only when the sensor works in the
passive mode because the sensor latency is measurable. 
NID is not suitable for a suspension attack
as such an attack model manages to suspend the sensor at the first attempt,
which hardly leaves any hint for NID to take effect.
Comparatively, Shade, as a hybrid detection method that
 closely collaborates with the robot control
 program, is not hindered by the working mode of sensors or attack models.

A notable observation revealed by~\autoref{tab:detect}
 is that the average reaction time of Shade
is much shorter or comparable than other detection
methods. For suspension attacks,  
CRV could detect them as well.
Given a suspension attack initiated at the startup of robot, 
only when the robot reached the safe distance (20cm) 
would CRV find that the sensor did not raise a `True' warning. 
This is why the average reaction time for CRV and Shade
is about 27s.
For fabrication attacks, fingerprinting could instantly detect
them. Meanwhile, the reaction
time of CRV is much shorter for fabrication 
attack model than two preceding attack models. It is 
because of 
the passive sensor mode with fabrication attacks.
Once the robot control program obtains a 
sensor value,
it asks CRV to check the numeric distance, which facilitates CRV compared to
boolean values used in the preceding two attack models.

\begin{figure}[t]	
	\centering
	\subfigure[Cleaned Distance]{\includegraphics[width=0.48\columnwidth]{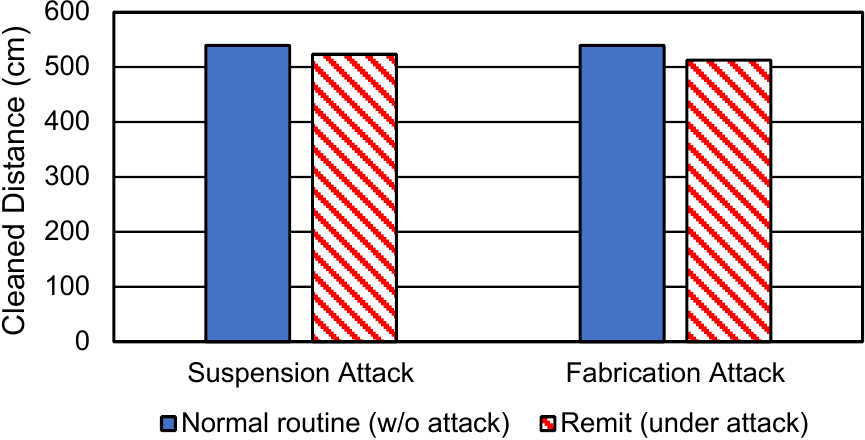}~\label{fig:mdist}}
	\hfill	
	\subfigure[Running time]{\includegraphics[width=0.48\columnwidth]{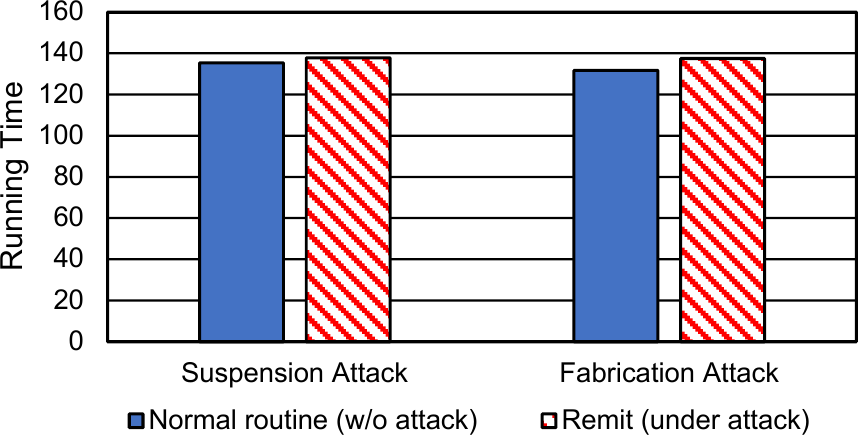}~\label{fig:mtime}}
	\caption{A Comparison between Remit with Attacks and Normal Routine}
	\label{fig:mit}	
\end{figure}

The default position to initiate an attack is when the robot starts up, so the attack is issued at a distance of 200cm to the obstacle. To further verify the efficiency of Shade, we did more tests when 
\RT triggered attacks at eight different distances to the left wall.
~\autoref{fig:range} captures four curves of reaction time for Shade.
In particular, given an attack occurring at a very short distance to the wall, 
say 25cm in~\autoref{fig:range}, Shade manages to detect it
at 0.6s to 3s, which efficiently protects the robot from security threats.

\subsection{Mitigation Results of Remit}

We have also done experiments to evaluate Remit.
The measurement of its effectiveness is the distance cleaned by the robot while its
efficiency is measured in terms of running time to clean the use case in Figure~\autoref{fig:case}.
We first made the robot clean the use case in a normal routine 
without any attack and recorded the cleaned distance as well as running time. 
Then, we 
ran Remit with the robot under attacks.
Figure~\ref{fig:mdist} and Figure~\ref{fig:mtime} present the 
results of cleaned distance and running time, respectively, for the normal routine and Remit.
Since Remit leverages the historical records maintained by Shade for cross-checking,
it can navigate the robot although the distance sensor is no longer reliable.
Owing to the accuracy limitation of records in navigation, Remit made losses of
3.3\% and 4.9\%, respectively, with two attack models. The overall loss 
is 4.1\%. Such insignificant losses confirm the effectiveness of Remit. 
On the other hand, after the robot entered the mitigation mode, Remit reduced the 
velocity of robot by 10\%. Though, as the robot cleaned 4.1\% less distances
under attacks, the total running time at the mitigation mode is eventually 
5.0\% more than that of the normal routine. 
To sum up, Remit not only accomplishes scheduled tasks but also
restricts extra time cost to an acceptable extent.

\section{Related Work}\label{sec:related}

CPS must be highly secure, especially for autonomous robotics systems~\cite{SABALIAUSKAITE2017174,robot:security:IntechOpen-2017,security:CPS:DATE-2017,ahmad2018analyzing}.
Researchers investigated the 
cyber threats to teleoperated surgical robots~\cite{robot:surgical-robot:2015,7579758}. For the cyber-security of industrial robots,
Quarta et al.~\cite{robot:industrial-security:SP-2017} performed a thorough 
analysis.
Comparatively, service robots are close to human beings, usually working together for service tasks~\cite{robot:service:RAS-2017}. Recently,
Lera et al.~\cite{robot:security:IntechOpen-2017} looked into the security threats with
a survey on the cyber-attacks associated to service robots as well as a
taxonomy that classifies the risks in using service robots.
However, not much work has been done to compromise  a movable service robot with
rational but harmful sensor values as \RT does. In particular,
Sabaliauskaite et al.~\cite{SABALIAUSKAITE2017174} comprehensively developed methods to conduct cyber-attacks to a  specific mobile
 robot. Whereas, their methods were significantly different from \RT since they tried to use irrational sensor values
 to crash the robot.

On the other hand,
how to detect and mitigate attacks for various CPS has been investigated~\cite{CPS:detection-survey:CSUR-2014,CPS:cross-layer:TCAD-2016,security:CPS:DATE-2017,fingerprint:CIDS:Security-2016,CPS:AM:WOOT-2017}.
For example,
Liu et al.~\cite{CSP:detection:TDSC-2016} used partially observable
Markov decision process to monitor and protect a
smart home against pricing attacks.
Dutta et al.~\cite{security:CPS-sensor:DAC-2017} utilized the
challenge response authentication to detect attacks
for active sensors and the recursive least square algorithm
to mitigate the impact of attacks.
Chhetri et al.~\cite{CPS:KCAD:ICCAD-2016} studied how to
detect an attack that could happen at various
points of the digital process chain of analog emissions in 
CPS like a 3D printer.

Researchers have also looked into security issues of service robots
in other aspects. For instance,
Guerrero-Higueras et al.~\cite{robot:localization-detection:RAS-2018} attended
attacks to real time location systems 
 for autonomous mobile robots.
Li et al.~\cite{robot:heavy-duty-robot:appl-sci} proposed to upload
the analysis of attack detection and mitigation to a cloud server in
the improved deep belief networks.
Our Shade and Remit differ from aforementioned approaches
in that they detect attacks within the 
computational resources of an autonomous service robot
and, furthermore, mitigate attacks without badly losing
the robot's work efficacy.

\section{Threats to Validity}~\label{sec:threats}

In this paper, we focus on protecting movable autonomous service robots. We use \RT to
fuzz sensor values that would impact the physical movement of robot. 
We leverage the historical records of obstacles to detect fuzzed sensor values
and navigate the robot to retain work efficacy. The limits of our proposals are
twofold. First, they are not directly applicable  
to non-movable autonomous robots.
Second, \RT fuzzes sensor values which are related to the movement of a robot; therefore, 
\RT does not cover how
to fuzz values for other types of sensors, e.g., the detectors for dust and water.

The two attack models considered in this paper, i.e., suspension attack and fabrication 
attack, are comprehensive and representative. Adversaries have managed to conduct such attacks
to CPS~\cite{fingerprint:CIDS:Security-2016}. 
These two attack models target compromising the sensor values and subsequently
misguide the robot control program. However, there exist
other attack models that are not discussed in this paper.
For example, a strong attacker may inject a malware in the control program; consequently,
the attacker does not rely on forging or suspending sensor values to manipulate the robot.

The Shade and Remit schemes which detect and mitigate attacks launched by \RT
demand the support of historical records of the environment. 
Thus, 
if a movable robot is placed in a fresh environment,
or new furnitures are installed in the original environment,
Shade and Remit might not function 
effectively as the records of such changed environments have not been fully obtained
 yet.

\section{Conclusion}~\label{sec:conclusion}

In this paper, we have considered security threats 
for autonomous service robots in order to protect them.
At the standpoint of developers, 
we propose \RT that automatically 
performs directed fuzzing in line 
with the normal state transitions of robot and the
environment where the robot works. 
By fuzzing sensor values at appropriate occasions,
\RT misleads the robot
to a rational but dangerous state so as to compromise it.

Moving even further, we develop Shade and Remit to detect and mitigate attacks
initiated through \RT, respectively. Shade and Remit
take advantage of historical records 
of obstacles to detect inconsistent obstacle appearances regarding untrustworthy sensor values
and navigate the movable service robot to  
continue working in motion.
As a result, we are able to 
efficiently detect and mitigate attacks 
but also retain the robot's work efficacy, which in turn enhances the security
and stability of autonomous service robot.
Experiments with a real-world cleaning robot show
that, 1) \RT dramatically outperforms fuzzing robot control program than state-of-the-art fuzzing tools, with much higher 
success rates of compromising the robot,
and 2) Shade and Remit maintain a high work efficacy at the mitigation mode with an insignificant loss.

\bibliographystyle{unsrt}
\bibliography{robot}

\begin{IEEEbiography}[{\includegraphics[width=1in,height=1.2in,clip,keepaspectratio]{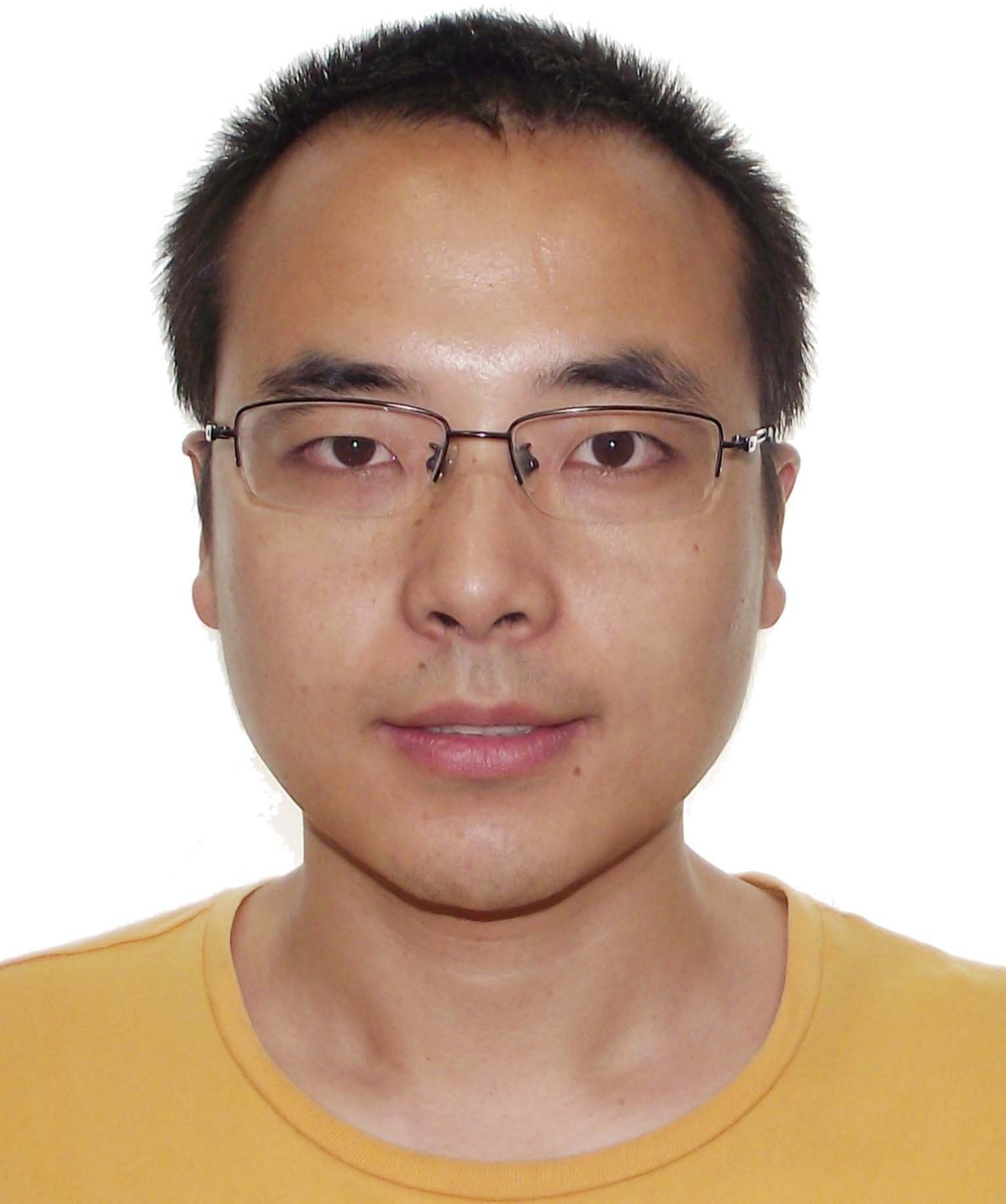}}]{Chundong Wang}
	received the Bachelor's degree in computer science from Xi'an Jiaotong University (2004-2008), and the Ph.D. degree in computer science
	from National University of Singapore (2008-2013). Currently he is
	a research fellow in Singapore University of Technology and Design (SUTD), Singapore.
	Before joining SUTD, he worked as a scientist in Data Storage Institute, A$^\star$STAR, Singapore.
	Chundong has published a number of papers in IEEE TC, ACM TOS, DAC, DATE, LCTES, USENIX FAST, etc.
	His research interests include data storage systems,
	non-volatile memory and computer architecture.
\end{IEEEbiography}

\begin{IEEEbiography}[{\includegraphics[width=1in,height=1.2in,clip,keepaspectratio]{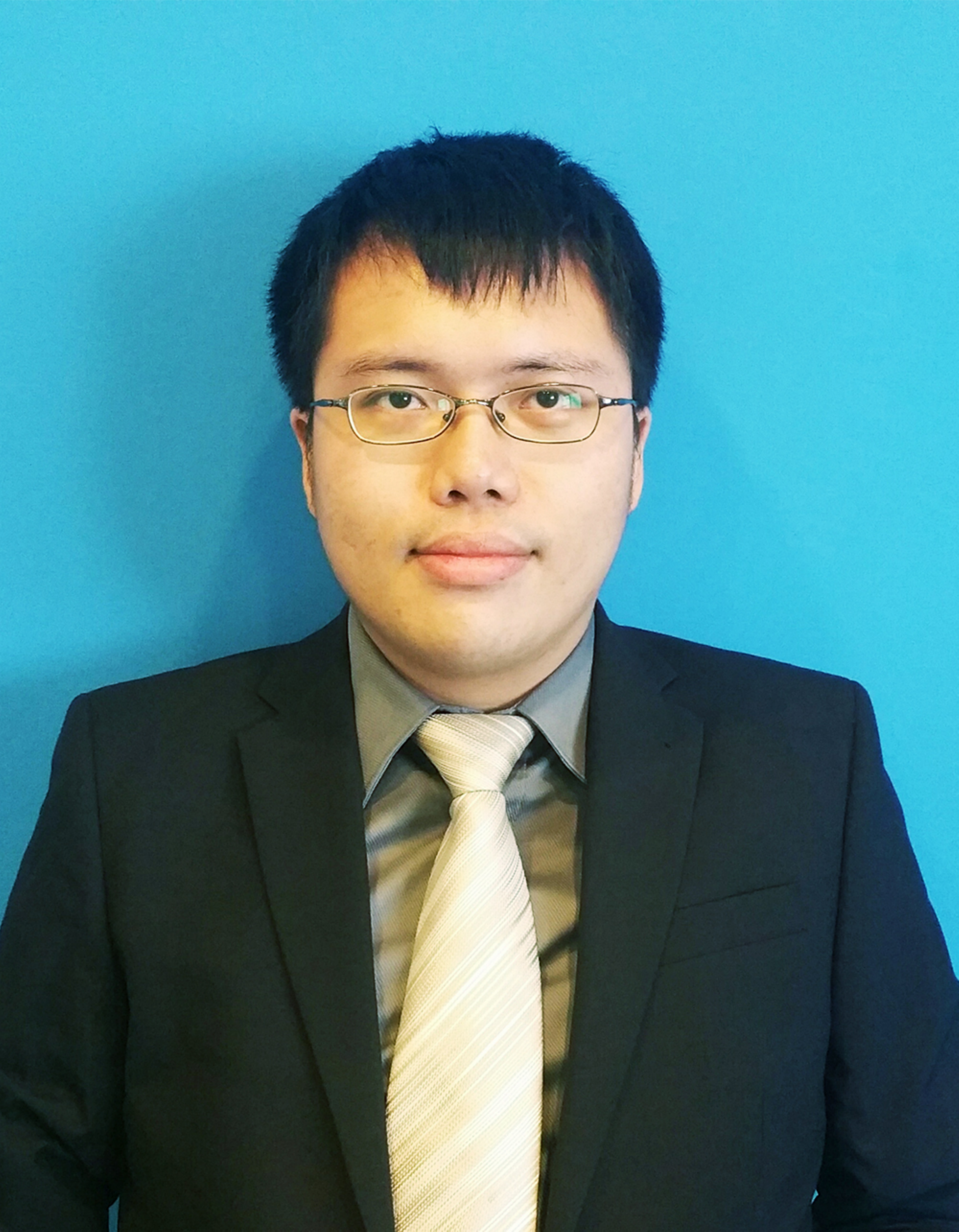}}]{Yee Ching Tok} received the Master's degree in information security from Royal Holloway, University of London, United Kingdom, in 2017. He is currently a PhD Student with the Information Systems Technology and Design Pillar, Singapore University of Technology and Design, Singapore. His current research interests are assessment of security in cyber-physical Systems, attack detection, and malicious software. Before pursuing his PhD degree, he worked as a Threat Hunter at Countercept where he helped to detect, respond and reduce impacts caused by malicious attackers to clients' critical assets. He has also carried out responsible disclosure of vulnerabilities to device manufacturers in his course of professional and academic research activities.

Yee Ching serves as an executive committee member in the Association of Information Security Professionals in Singapore.
\end{IEEEbiography}

\begin{IEEEbiography}[{\includegraphics[width=1in,height=1.2in,clip,keepaspectratio]{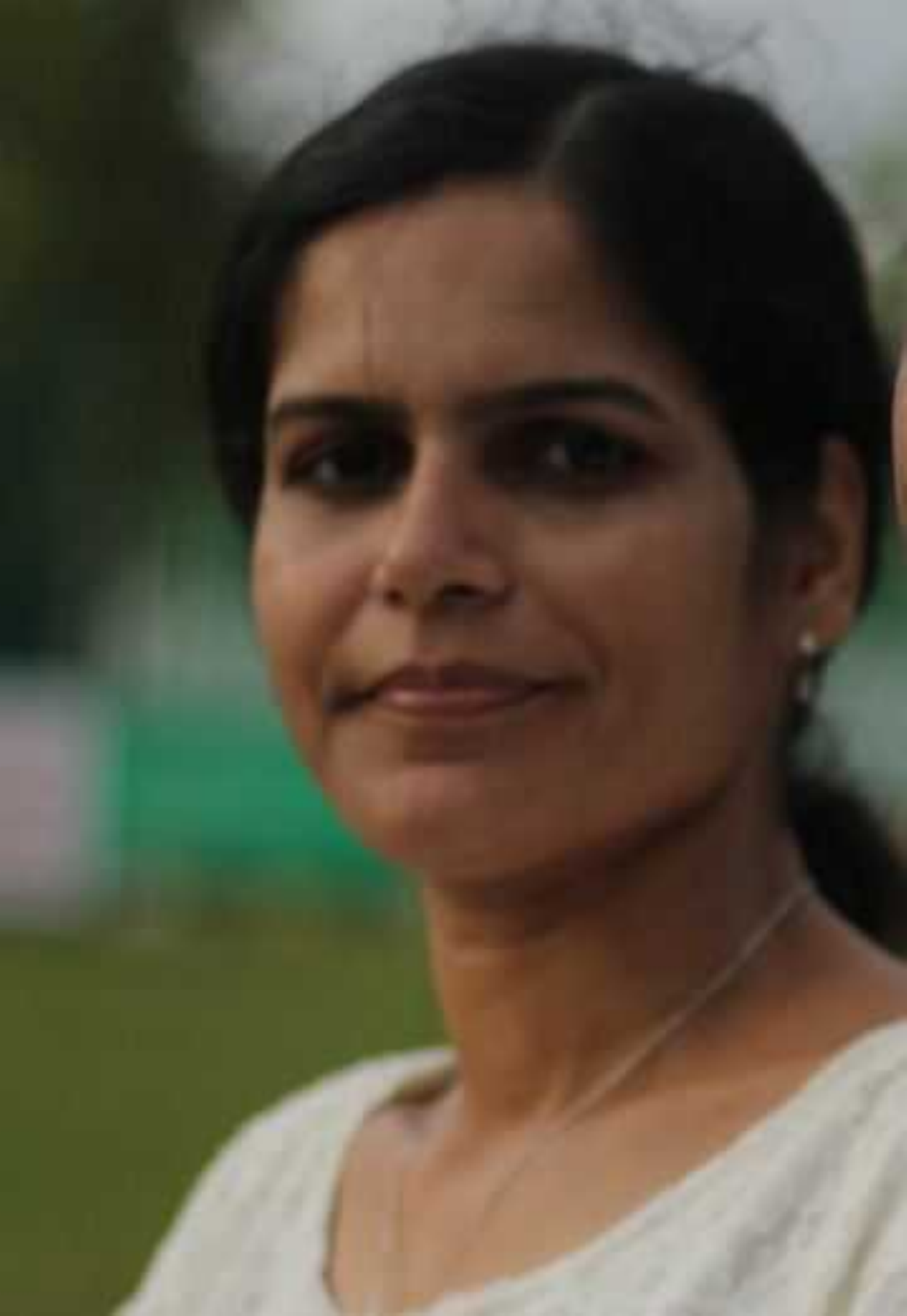}}]{Rohini Poolat} received the Master of Technology in Software Engineering from the National University of Singapore, Singapore in 2009. She is a research assistant in the cyber security research team at Singapore University of Technology and Design (SUTD), Singapore. She has worked on all phases of the software development cycle in various industry projects before moving into the research area. Her research interests include cyber-attack detection, mitigation and solutions to prevent attacks.
\end{IEEEbiography}

\begin{IEEEbiography}[{\includegraphics[width=1in,height=1.2in,clip,keepaspectratio]{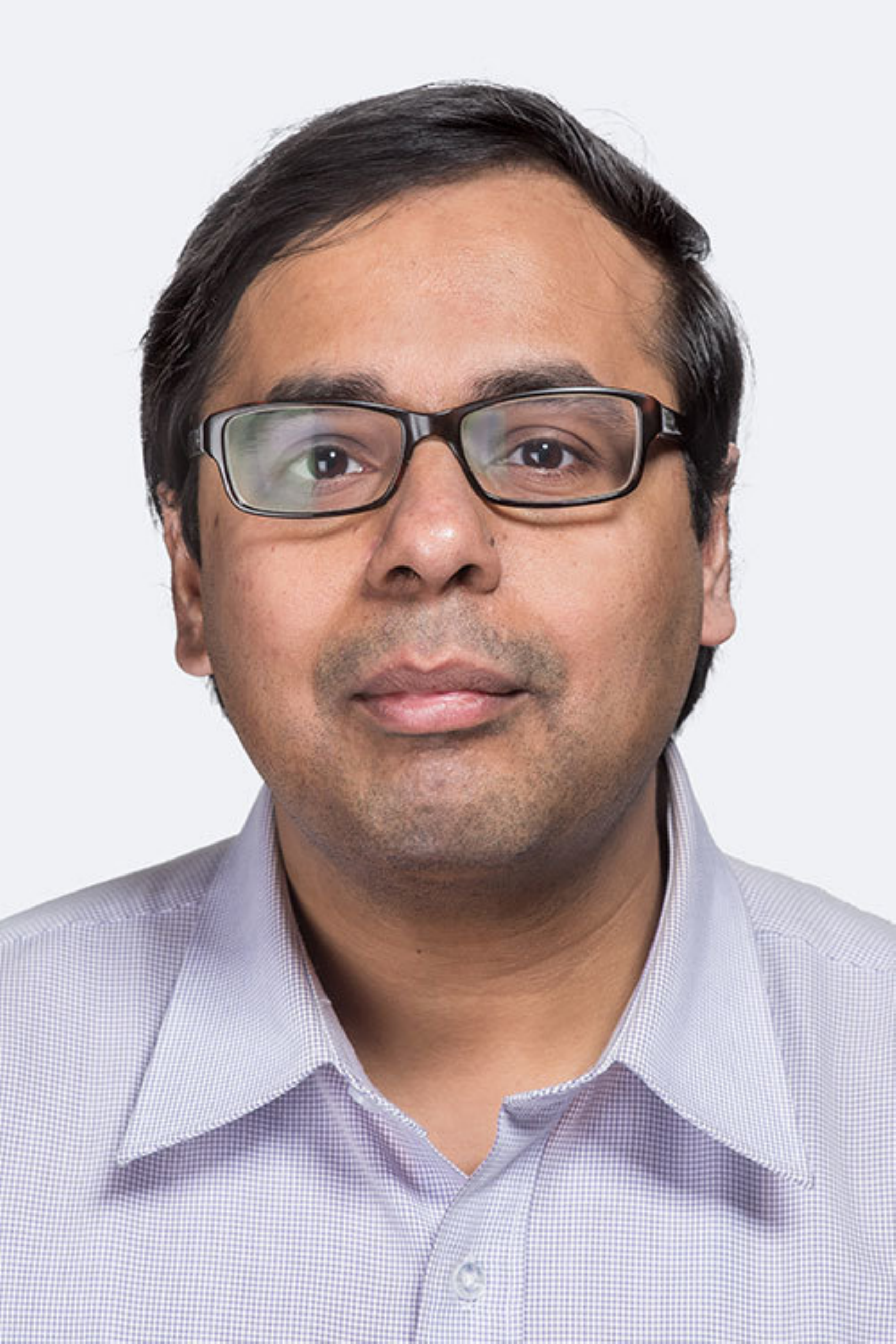}}]{Sudipta Chattopadhyay} received the Ph.D. degree in computer science from the National University of Singapore, Singapore, in 2013.
	 He is an Assistant Professor with the Information Systems Technology and Design Pillar, Singapore University of Technology and Design, Singapore. In his doctoral dissertation, he researched on Execution-Time Predictability, focusing 
	 on Multicore Platforms. He seeks to understand the influence of execution platform on critical software properties, such as performance, energy, robustness, and security. His research interests include  program analysis, embedded 
	 systems, and compilers.
	
	Mr. Chattopadhyay serves in the review board of the IEEE Transactions on Software Engineering.
\end{IEEEbiography}

\begin{IEEEbiography}[{\includegraphics[width=1in,height=1.2in,clip,keepaspectratio]{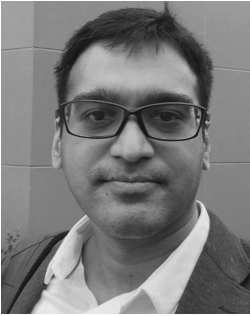}}]{Mohan Rajesh Elara}
received the B.E.
degree from the Bharathiar University, India,
in 2003, and the M.Sc. and Ph.D. degrees from
Nanyang Technological University in 2005 and
2012, respectively. He is currently an Assistant
Professor with the Engineering Product Development 
Pillar, Singapore University of Technology
and Design. He is also a Visiting Faculty Member
with the International Design Institute, Zhejiang
University, China. He has published over 80 papers
in leading journals, books, and conferences. His research interests are
in robotics with an emphasis on self-reconfigurable platforms as well as
research problems related to robot ergonomics and autonomous systems.
He was a recipient of the SG Mark Design Award in 2016 and 2017,
the ASEE Best of Design in Engineering Award in 2012, and the Tan Kah
Kee Young Inventors’ Award in 2010.
\end{IEEEbiography}

\end{document}